\documentclass[acmtog,screen]{acmart}
\AtBeginDocument{%
  }

\newcommand{\dproj}{\mathrm{d} \sigma_\bot}
\newcommand{\hemi}{\Omega^+}
\newcommand{\fdot}[2]{\langle #1, #2 \rangle}

\def\figurePath{fig/}
\def\myfigure#1#2#3{\begin{figure}[tb]\centering\includegraphics[width = \linewidth]{\figurePath#2}\caption{#3}\label{fig:#1}\end{figure}}

\newcommand{\added}[1]{{\color{black}#1}}

\newcommand{\revise}[1]{\textcolor{black}{#1}}

\copyrightyear{2026}
\acmYear{2026}
\setcopyright{cc}
\setcctype{by}
\acmConference[SIGGRAPH Conference Papers '26]{Special Interest Group on Computer Graphics and Interactive Techniques Conference Conference Papers}{July 19--23, 2026}{Los Angeles, CA, USA}
\acmBooktitle{Special Interest Group on Computer Graphics and Interactive Techniques Conference Conference Papers (SIGGRAPH Conference Papers '26), July 19--23, 2026, Los Angeles, CA, USA}
\acmDOI{10.1145/3799902.3811156}
\acmISBN{979-8-4007-2554-8/2026/07}

\begin{document}

\title{PureSample: Neural Materials Learned by Sampling Microgeometry}

\author{Zixuan Li}
\orcid{0009-0004-2424-9529}
\affiliation{
    \institution{College of Computer Science, Nankai University}
    \city{Tianjin}
    \country{China}
}
\email{zixuan.li_2001@outlook.com}

\author{Zixiong Wang}
\orcid{0000-0002-6170-7339}
\affiliation{
    \institution{College of Computer Science, Nankai University}
    \city{Tianjin}
    \country{China}
}
\email{zixiong_wang@outlook.com}

\author{Jian Yang}
\orcid{0000-0003-4800-832X}
\affiliation{
    \institution{Nankai University and Nanjing University}
    \city{Tianjin}
    \country{China}
}
\email{csjyang@nankai.edu.cn}

\author{Milo\v{s} Ha\v{s}an}
\orcid{0000-0003-3808-6092}
\affiliation{
    \institution{NVIDIA}
    \city{San Jose}
    \country{USA}
}
\email{milos.hasan@gmail.com}

\author{Beibei Wang}
\orcid{0000-0001-8943-8364}
\authornote{Corresponding author.}
\affiliation{
    \institution{School of Intelligence Science and Technology, Nanjing University}
    \city{Suzhou}
    \country{China}
}
\email{beibei.wang@nju.edu.cn}
\begin{abstract}
    Traditional physically-based material models rely on analytically derived bidirectional reflectance distribution functions (BRDFs), typically by considering statistics of micro-primitives such as facets, flakes, or spheres, sometimes combined with multi-bounce interactions such as layering and multiple scattering. These derivations are often complex and model-specific. Once an analytic BRDF evaluation is defined, one still needs to design an importance sampling method for it and evaluate the probability density function (pdf) of that sampling distribution, requiring further model-specific derivations.

    We present \emph{PureSample}: a novel neural BRDF representation that allows learning a material's appearance purely by sampling forward random walks on the microgeometry, which is usually straightforward to implement. Our representation allows for efficient BRDF evaluation, importance sampling, and pdf evaluation, for homogeneous as well as spatially varying materials.
    We achieve this by two learnable components: first, the sampling distribution is modeled using a flow matching neural network, which allows both importance sampling and pdf evaluation; second, we introduce a view-dependent albedo term, captured by a lightweight neural network, which allows for converting a pdf value to a BRDF value for any pair of view and light directions.
    We demonstrate PureSample on challenging materials, including various microgeometries, multi-layered materials, and multiple-scattering microfacet materials.
    
\end{abstract}




\begin{teaserfigure}
\centering
\includegraphics[width=\textwidth]{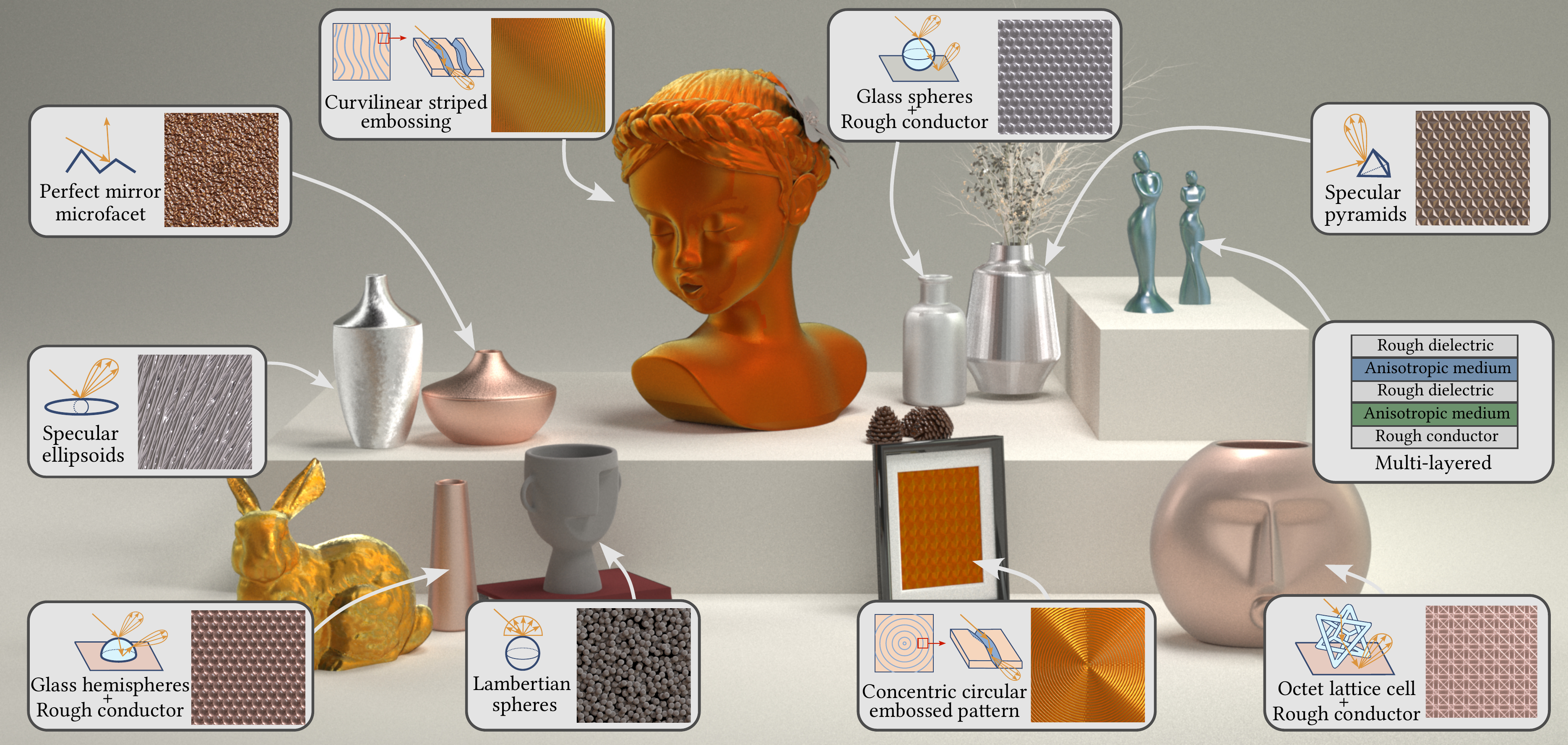}
\caption{We present PureSample, a novel neural BRDF representation that allows learning a material's appearance purely by sampling forward random walks on an arbitrary microgeometry. We showcase a variety of microgeometries within a single scene, including different micro-primitives and layered materials, highlighting PureSample's expressive capabilities. A boxed inset next to each rendering visualizes the corresponding microgeometry.}
\label{fig:teaser}
\end{teaserfigure}

\maketitle

\section{Introduction}
\label{sec:intro}

\begin{figure*}[!t]
\centering
\includegraphics[width = 0.87\linewidth]{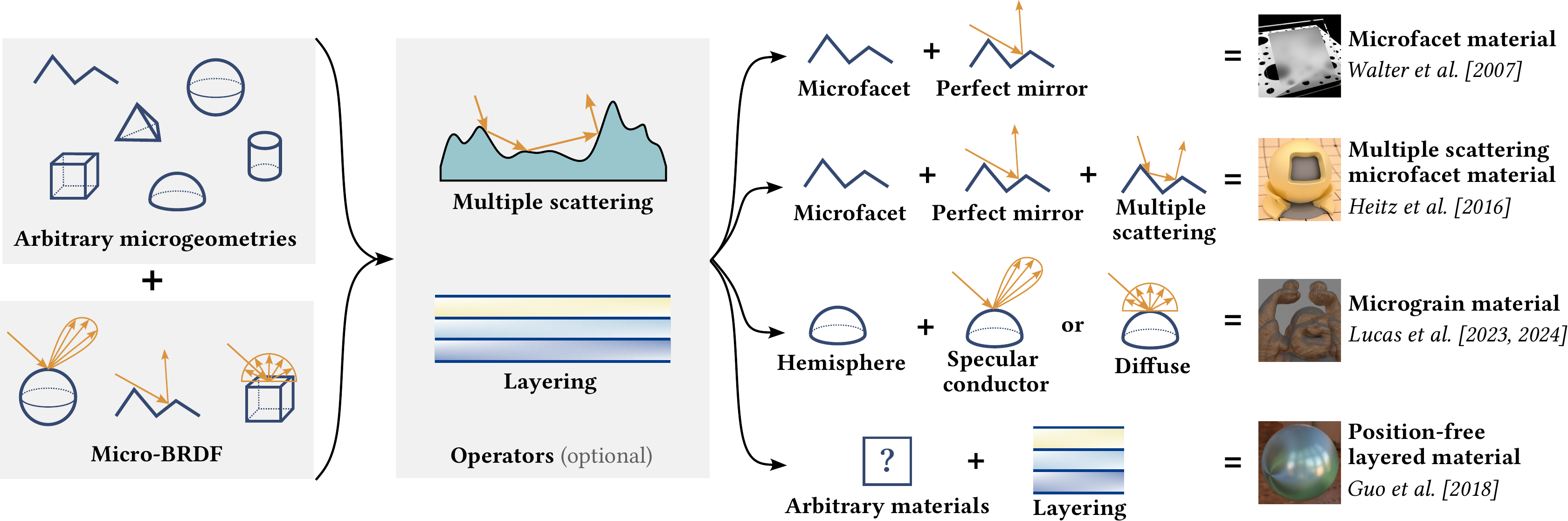}
\caption{In typical physically-based material models, a material consists of three components: a microgeometry, a micro-BRDF, and an ``operation'' (multiple bounce, layering, etc.). Our framework not only unifies these models but also allows extending the material models by introducing arbitrary microgeometries.}
\label{fig:complexBRDF}
\end{figure*}

Computer graphics has been successful at approximating real-world appearances using bidirectional reflectance distribution functions (BRDFs), which describe how surfaces respond to light reflection. The field has developed numerous analytical BRDFs, ranging from models based on microfacets ~\cite{Cook:1982:ReflectanceModel, Walter:2007:Microfacet} and microflakes ~\cite{Jakob:2010:Microflake,Heitz:2015:SGGX,Wang:2023:SpongeCake}, to ones based on sphere microgeometries ~\cite{dEon:2024:VMFDiffuse} or hemisphere microgeometries ~\cite{Lucas:2023:micrograin}. These models can be categorized by their structure, such as single-layer versus multi-layer ~\cite{Belcour:2018:Layered, Guo:2018:layered, Wang:2023:SpongeCake}, or by their interactions with light, including single bounce versus multiple bounces \cite{Bitterli:2022:pfmultiMicrofacet,Cui:2023:GMBBRDF,Heitz:2016:multimicrofacet,Wang:2022:MBBRDF}. However, these models tend to have complex derivations that are largely model-specific; organizing them into a cohesive framework, simplifying their derivations, and extending the set with new models remains a challenge.



Most current physically-based BRDF models essentially consist of three components: a microgeometry, a micro-BRDF, and an ``operation'', as illustrated in Fig.~\ref{fig:complexBRDF}. The microgeometries could include microfacets, microflakes, microspheres, etc. On top of these, a micro-BRDF describes the light interaction with an isolated microgeometry, assuming rules of geometric optics still apply at the given scale. For example, applying a perfectly specular (mirror) micro-BRDF to a microfacet, and considering a single bounce leads to the standard microfacet model, while a Lambertian micro-BRDF applied to microspheres~\cite{dEon:2024:VMFDiffuse} produces a different model for rough surfaces. Finally, operations such as multiple bounces or layering can enhance the expressive capabilities of these models.

All of these models require specific non-trivial techniques to derive their statistical aggregate formulations. Introducing a new microgeometry requires significant effort, which becomes even more challenging when considering multiple bounces, multiple different microgeometries, layering among the microgeometries, or spatial variation. This is because the value of a BRDF is essentially a path integral over all microgeometry paths that connect the incoming-outgoing ray pair, and there are no standard techniques to derive the closed-form solution for such integrals. This complexity in deriving new BRDF has hindered the development of innovative models. Once an analytic BRDF's evaluation is defined, one still needs to design an importance sampling method for it, and a way to evaluate the probability density function~(pdf) of that sampling distribution, requiring further model-specific derivations. Therefore, simplifying the derivation of new models is becoming essential. 

Neural networks have been introduced to represent and render complex materials, including layered materials~\cite{Fan:2022:Neurallayer,Guo:2023:metalayer}, multi-bounce microfacet models~\cite{Xie:2019:MultipleScatteringML}, and synthetic or measured mesostructures~\cite{sztrajman2021nbrdf, Kuznetsov:2021:neumip, Kuznetsov:2022:NeuralMatCurve}. Importance sampling can be added to the final models~\cite{Xu:2023:NeuSample, Wu:2025:Reparam, Fu:2024:BSDFdiffusion}, but a ground truth BRDF formulation is still necessary for rendering the synthetic data to which these networks are fitted. This leads to an interesting question: Is it possible to eliminate the BRDF derivation entirely, and learn a neural material directly from a microgeometry?

In this paper, we present a novel neural framework called \emph{PureSample}, designed to learn any material described by an explicit microgeometry. The key insight is that sampling a material’s behavior through forward particle tracing (i.e., finding the outgoing direction of a random walk for a given incoming direction) is easy to implement for most microgeometries, simply by importance sampling the micro-BRDF at each particle interaction with a micro-primitive until exiting the surface (or being absorbed). The sampled directions produce an outgoing distribution tailored to the given incoming direction. Note that no next-event estimation~(NEE) or bidirectional tracing is needed, which is an important advantage, since these can be non-trivial to implement even for specific restricted microgeometries \cite{Yan:2014:Glints,Guo:2018:layered}.

Leveraging recent progress in neural sampling models, we show that the conditional outgoing distribution can be efficiently learned from the sampled directions. 
While there is no direct way to extract BRDF values from the samples of forward path tracing, the corresponding albedo can be readily obtained, allowing us to construct a BRDF evaluation as the product of two components: 
a probability distribution term and a view-dependent albedo term, where the former models the overall normalized shape of the reflectance function and the latter models the energy absorption and color variation across different channels. In practice, the distribution term is represented by a flow matching model, and the albedo term is represented by a small multi-layer perceptron (MLP). Both terms are learned purely from the sample distributions generated by particle tracing random walks on the microgeometry, hence the method name.

During rendering, the learned distribution and the albedo network work together to provide all the necessary operations of a material system (BRDF evaluation, importance sampling, and pdf evaluation). We validate the PureSample framework on several microgeometries, including spheres, ellipsoids, pyramids, facets, layers, embossing, and more, including multiple bounces. The resulting renderings match their corresponding existing models (when available). For all these materials, our framework successfully replicates the appearance of the ground truth microgeometries. Consequently, the PureSample framework can serve as a universal way to represent the material implied by an arbitrary microgeometry as an appearance model (as long as a forward path tracing implementation is available for it, which is generally straightforward),  and easily apply it to arbitrary meshes.
%
To summarize, our main contributions include: 
\begin{itemize}
  \item a novel BRDF representation framework, \emph{PureSample}, capable of representing a wide range of materials purely from sampled light paths of an explicit microgeometry, with no need to derive BRDF formulations,
  \item a neural architecture that uses flow matching and a view-dependent albedo term to achieve all three necessary operations of a material system (BRDF evaluation, importance sampling, and pdf evaluation) for this representation, including support for spatial variation, and
  \item a demonstration of the effectiveness of PureSample on a wide range of materials, including classically studied microgeometries where a previously derived BRDF is available, as well as more complex novel cases.
\end{itemize}
\section{Related work}
\label{sec:relatedwork}
In this section, we briefly review related work on physically-based appearance models, neural BRDF representations, and neural importance sampling approaches. 

\subsection{Physically-based appearance models}
We focus on appearance models under geometric optics.
Microfacet materials~\cite{Cook:1982:ReflectanceModel,Walter:2007:Microfacet}, often combined with a diffuse term, are widely used to represent a variety of appearances, ranging from metals to plastics and from smooth to rough surfaces. Historically, these models handled only single scattering from the microfacets. They have been later extended to handle multiple scattering~\cite{Heitz:2016:multimicrofacet,Bitterli:2022:pfmultiMicrofacet,Wang:2022:MBBRDF,Cui:2023:GMBBRDF} to ensure energy conservation. Ribardière et al.~\shortcite{Ribardire:2019:NDFMicrogeo} use a virtual gonioreflectometer to analyze the effect of different NDFs on microfacet materials. Recently, d’Eon et al.~\shortcite{d'Eon:2023:Non-Uniform} introduced an extension for modeling microfacet materials with a normal distribution that varies with elevation.

Other microgeometries have been introduced into appearance modeling. Spheres have been used with specular mirror~\cite{Chandrasekhar1960} or Lambertian reflection~\cite{dEon:2024:VMFDiffuse}. More recently, the hemisphere has been introduced to model the top layer as a micrograin model~\cite{Lucas:2023:micrograin,Lucas:2024:anisoMicrograin}, enabling the rendering of effects such as dust, dirt, and spray paint. 

Current approaches for layered materials can be classified into several categories: approximate analytic models \cite{Weidlich:2007:layering}, Fourier basis functions \cite{jakob:2014:layered,Jakob2015Layerlab,Zeltner2018Layer}, tracking directional statistics \cite{Belcour:2018:Layered,Yamaguchi:2019:anisotropic,WeierAndBelcour:2020:Anisotropic}, Monte Carlo simulation-based methods with next-event estimation or bidirectional tracing \cite{Guo:2018:layered,Gamboa:2020:EfficientLayered,Xia:2020:Layered}, and an approach that maps multiple scattering among layers to a single scattering lobe using neural networks~\cite{Wang:2023:SpongeCake}. A closely related line of work is bi-scale material design~\cite{Wu:2011:biscaleDesign,Wu:2013:inverseBiscale}. These approaches generate diverse large-scale materials by designing small-scale geometry. However, they approximate the original small-scale geometry using facets, and cannot fully capture the appearance of arbitrary microgeometries.

Some of these physically-based appearance models rely on carefully crafted assumptions and derivations, while others require Monte Carlo sampling with next-event estimation or bidirectional tracing. In contrast, our method only needs forward path tracing to generate sampled directions, allowing for arbitrary microgeometries. This significantly simplifies both material design and evaluation.


\subsection{Neural material representations}

Neural networks have been introduced into BRDF modeling due to their ability to compress data, accelerate rendering, and represent effects that are challenging to model analytically. Several works~\cite{Fan:2022:Neurallayer,Guo:2023:metalayer,Zeltner:2024:RealTimeAppearance} represent layered materials using different types of networks to avoid random walks within layered materials, thereby reducing both rendering and storage costs. Kuznetsov et al.~\shortcite{Kuznetsov:2019:learnMicrogeo} try to learn NDF from microgeometries, which use a generative adversarial network to capture specular highlights. Besides complex synthetic materials, neural networks have also been utilized to represent materials for compressing~\cite{Rainer2019Neural,Rainer2020Unified,Hu:2020:DeepBRDF,sztrajman2021nbrdf,Fan:2023:BTF,Shi:2022:LearnBiScale,TG:2024:NeuralSSS,Zheng:2021:NeuralProcessBRDF,Dou:2024:SphericallyBRDF,tg:2024:neupress} or synthesis~\cite{Xu:2024:BTF,Xu:2025:towards}. In addition to single-scale representation, neural BRDF models~\cite{Kuznetsov:2021:neumip,Kuznetsov:2022:NeuralMatCurve} also allow multiscale representation. 

While these neural material models can avoid complex evaluation during rendering (e.g., a random walk in the medium or the microgeometries), the BRDF formulation is still required for training these networks on synthetic data. In contrast, our method can eliminate the BRDF derivation entirely in both training and rendering. 


\begin{figure*}[!t]
\centering
\includegraphics[width = 0.89\linewidth]{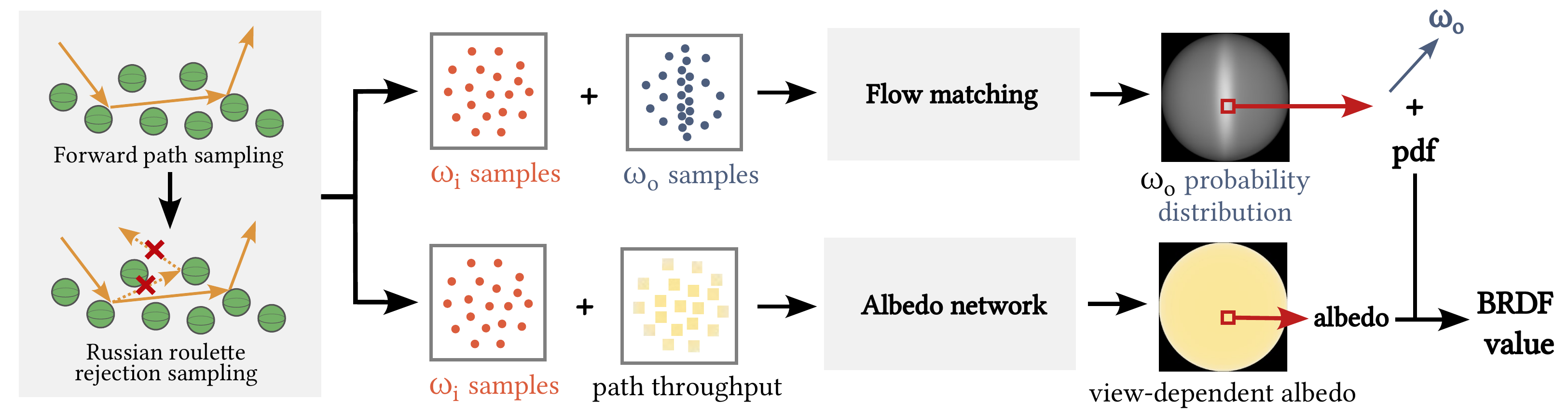}
\caption{Overview of our PureSample framework. For a given microgeometry, we sample the incoming direction uniformly and perform a forward path sampling with Russian roulette rejection sampling, generating outgoing directions with path throughput~(sampling weights). With these samples, we train a flow matching model for the probability distribution term (conditioned on $\omega_i $ and color channel) and an MLP for the view-dependent albedo term. With these two networks, our PureSample representation allows BRDF evaluation, importance sampling, and pdf evaluation.}
\label{fig:pipeline}
\end{figure*}

\subsection{Neural BRDF importance sampling}
Beyond BRDF representation, neural networks have been utilized for BRDF importance sampling. These approaches can generally be classified into three categories: fitting analytic lobes or mixtures~\cite{sztrajman2021nbrdf, Fan:2022:Neurallayer}, learning the probability distribution using neural models such as normalizing flows~\cite{muller:2019:neuralis, Litalien:2024:Warp}, and predicting an approximate discrete distribution (histogram). All of these methods are studied and improved by Xu et al.~\shortcite{Xu:2023:NeuSample}.

Related to our work are NeuSample~\cite{Xu:2023:NeuSample} and the diffusion BSDF sampling method by Fu et al.~\shortcite{Fu:2024:BSDFdiffusion}. NeuSample derives importance sampling for a neural material by sampling from the tabulated reflectance and fitting analytic lobes or normalizing flows to the samples, conditional on a feature texture for spatial variation. Fu et al.~\shortcite{Fu:2024:BSDFdiffusion} show that flow matching~\cite{lipman2023flowmatchinggenerativemodeling} is effective for importance sampling but requires multi-step sampling, which limits rendering performance.
Recently, Wu et al.~\shortcite{Wu:2025:Reparam} introduced a reparameterization-based method for neural BRDF importance sampling, which enables accurate and efficient BRDF sampling. However, their network is nontrivial to invert, which poses a challenge for evaluating the sampling pdf for arbitrary directions. All these methods assume a ground truth reflectance function already exists, and produce a sampling routine for it. \revise{Our contribution is to avoid the derivation of ground-truth reflectance} and demonstrating that the samples from a microgeometry are sufficient to learn the importance sampling pdf, and by further extensions, the entire spatially varying material.

\section{Motivation and overview}
\label{sec:system}



\added{Physically-based materials are typically modeled by the statistical behavior of a microgeometry (often but not always an assembly of micro-primitives). Examples include microfacet and microflake models~\cite{Jakob:2010:Microflake, Heitz:2015:SGGX}, as well as sphere- and hemisphere-based microgeometries~\cite{Chandrasekhar1960, dEon:2024:VMFDiffuse, Lucas:2023:micrograin}. Beyond these typical cases, microgeometry can comprise arbitrary slabs of reflective, refractive, or scattering surfaces and media. In Fig.~\ref{fig:complexBRDF}, we show several examples of previously studied microgeometries and several new ones.}

Built upon these microgeometries, different kinds of operations are typically considered, including multiple bounces, layering and volumetric scattering. Even if the single-scattering formulation has a closed-form solution, other components such as the multiple-scattering term or layering, typically require Monte Carlo estimation or specific numerical approximations. Our method transparently handles all of these cases, since the learned distribution and albedo term are based purely on the sampled data and completely agnostic to the microgeometry itself.


A key observation is that sampling a material behavior through forward particle tracing is easy to implement for most microgeometries, simply by importance sampling the micro-BRDF (or micro-phase function) at each particle interaction with the microgeometry and continuing tracing until exiting the surface. \added{As shown in Fig.~\ref{fig:pipeline},} the sampled directions define an outgoing pdf conditional on the given incoming direction; the pdf can be further conditioned on the spatial position of the material. This pdf can be efficiently learned by a neural model, leveraging recent advances in diffusion models and flow matching, and can be further converted into a BRDF by learning a view-dependent albedo term. Consequently, the representation and learning routine for BRDFs and spatially varying BRDFs (SVBRDFs) across many microgeometries becomes standardized.

\section{PureSample: BRDF decomposition}
\label{sec:formulation}
\begin{table}[h]
	\caption{\label{tab:notations} Notation summary.}
\begin{small}
  \begin{tabular}{|l|c|l|}\hline
		$\omega_i / \omega_o $ & incident / outgoing direction \\
		$f(\omega_i, \omega_o)$ & BRDF function \\
        $\rho(\omega_o | \omega_i)$ & conditional pdf term \\
		$\alpha(\omega_i)$ & view-dependent albedo term \\
        ${\Omega^+}$ & upper hemisphere \\
        \hline
        $c$ & flow matching condition ($\omega_i$, channel, feature) \\
        $g$ & normalized unit 2D Gaussian \\
        $\theta$ & parameters of flow matching network \\
        $u_{\theta}$ & flow velocity prediction \\
        $t$ & flow ``time'' parameter in $[0,1]$ \\
        \hline
		\end{tabular}

\end{small}
\end{table}

In this section, we present the theory of our PureSample framework for BRDF representation. For clarity, the symbols and terms used throughout our paper are summarized in Tab.~\ref{tab:notations}.

\subsection{Distribution/albedo decomposition}

A BRDF (without spatial variation) is a 4D function $f(\omega_i,\omega_o)$, \added{where $\omega_i$ denotes the incident (view) direction, and $\omega_o$ denotes the outgoing (light) direction. }To learn BRDFs from samples of microgeometries\added{, where direct BRDF evaluation is unavailable, we propose a decomposition of the BRDF as the product of two components}: a probability distribution term and a view-dependent albedo term, where the former models the overall normalized shape of the reflectance function and the latter models the energy absorption and color variation across different color channels.  

For simplicity, we first consider a BRDF that is homogeneous (without spatial variation) and scalar (grayscale). In this case, we can write an exact decomposition of the BRDF:  
\begin{equation}
    f(\omega_i,\omega_o) = \rho(\omega_o | \omega_i) \ \alpha(\omega_i),
    \label{eq:pure-sampling}
\end{equation}
where
\begin{equation}
    \begin{aligned}
    \rho(\omega_o | \omega_i) = \frac{f(\omega_i,\omega_o)}{\alpha(\omega_i)}, \ \ \alpha(\omega_i) = \int_{\hemi} f(\omega_i,\omega)\dproj(\omega).
    \end{aligned}
    \label{eq:ps_general}
\end{equation}
Here $\rho(\omega_o | \omega_i)$ can be intuitively seen as the probability distribution of a particle exiting the surface in direction $\omega_o$, given incident direction $\omega_i$, and given that the particle is not absorbed by the microgeometry. The second term $\alpha(\omega_i)$ can be seen as a view-dependent albedo term, representing the preserved energy of particles with incident direction $\omega_i$. Note that $\sigma_\bot(\omega)$ is the projected solid angle measure, and that $\rho$ is defined with respect to this measure, and can be equivalently seen as a 2D distribution on the projected hemisphere.
 The above decomposition exists for any reasonable scalar BRDF \revise{($f(\omega_i,\omega_o) \geq 0$ and the integral is finite and positive)}. In practice, we do not know $\rho$ or $\alpha$. However, given any microgeometry, we can estimate both terms using a Monte Carlo process.

\subsection{Monte Carlo estimator for the decomposition}
\label{sec:pure-sampling_gen}


Starting from an incident particle (ray) from direction $\omega_i$, we perform forward path tracing until the particle exits the surface. During this random walk, the particle interacts with surfaces or scatters within media. For both event types, we employ importance sampling on the micro-BRDF or the medium's phase function to determine the next ray direction. This importance sampling operation yields a weight equal to the ratio of the BRDF value at the sampled direction to the sampling pdf. We perform Russian roulette rejection sampling to discard the samples with probability equal to one minus the weight (note that we are still assuming a scalar BRDF, and therefore a scalar weight). If the path survives, it is divided by the probability of survival, which cancels the weight, and maintains a unit path ``throughput'' (i.e., the product of accumulated terms along the path). Once the ray exits the surface, we record the outgoing direction $\omega_o$. We also record the number of accepted (non-terminated) samples.

This way, we have defined a particle tracing Monte Carlo process, whose distribution of ray directions leaving the surface is precisely $\rho(\omega_o | \omega_i)$, and the expected value of the fraction of accepted samples is precisely $\alpha(\omega_i)$. For a more formal proof, please see the supplementary materials.

To turn this Monte Carlo process into a practical material representation, we still need to resolve the following:
\begin{itemize}
    \item We would like to learn the distribution $\rho$ and the albedo $\alpha$ from the sampled data, instead of running the Monte Carlo process during rendering.
    \item We need a way to \emph{evaluate} the pdf $\rho(\omega_o | \omega_i)$ for any incoming and outgoing direction, not just \emph{sample} from it.
    \item We would like to support BRDFs with color channels and spatially varying microgeometry.
\end{itemize}
These issues are addressed in the next section.


\section{PureSample: Neural representation}
\label{sec:representation}

Flow matching models have recently demonstrated a powerful capacity to represent conditional probability distributions and learn them from samples, while enabling both sampling and evaluation of the learned distribution, becoming a perfect choice for our task. We leverage flow matching for the probability distribution representation (see Sec.~\ref{sec:flow}) and a small MLP for the view-dependent albedo term (see Sec.~\ref{sec:albedo}). We add RGB color support by additionally making the probability distribution term conditional on the color channel, and using RGB output for the albedo network. Additionally, we introduce a latent (feature) texture to enable spatially varying materials (see Sec.~\ref{sec:sv}). Then, we integrate both networks for rendering (see Sec.~\ref{sec:rendering}). \added{Finally, we introduce our optional network acceleration strategies (see Sec.~\ref{sec:acceleration}).} 


\myfigure{FMNetwork}{flowmatchingNetwork.pdf}{We train our flow matching network using randomly sampled $t$ and corresponding $\omega_i,\omega_o$ samples. During inference, we invoke the flow matching network within the ODE solver to transform samples from the base distribution to the target distribution.}


\subsection{Flow matching for the probability distribution term}
\label{sec:flow}


The term \(\rho(\omega_o | \omega_i)\) describes the probability distribution of \(\omega_o\) conditional on \(\omega_i\). This distribution is defined on the projected hemisphere space (unit disk). We employ flow matching to predict a velocity field that transforms a base distribution $g$ into the target distribution \(\rho\). By applying the discretized flow field, we can effectively sample from the learned target distribution, yielding a sampled direction along with its pdf.


The flow matching utilizes an MLP denoted as \(u_{\theta}\), which consists of five layers (64 neurons per layer), as shown in Fig.~\ref{fig:FMNetwork}. The input condition \(c\) for the network includes the incoming direction along with a color channel condition represented by one-hot encoding. This channel condition allows the network to learn different distributions across RGB color channels. For the base distribution $g$, we use a 2D Gaussian distribution with a mean of 0 and a variance of 1.

Following previous works~\cite{lipman2023flowmatchinggenerativemodeling, Fu:2024:BSDFdiffusion}, we use the conditional flow matching (CFM) loss for training:
\begin{equation}
    \begin{aligned}
      \mathcal{L}_{\mathrm{CFM}}(\theta)&=\|u_{\theta}(x_{t}, c, t)-(x_{1}-x_{0})\|^{2},\\
        x_{t} &= tx_{1}+(1-t)x_{0},\\
        x_0 &\sim g,x_1\sim \rho,
    \end{aligned}
    \label{eq:FM_loss}
\end{equation}
\added{
where \(c\) represents the conditioning variable that combines \(\omega_i\) and the color channel. 
\(x_0\) and \(x_1\) denote samples drawn from the target distribution \(\rho\) and the base Gaussian distribution \(g\), respectively. 
For the time step \(t\), we sample from a mixture of two uniform distributions $\mathcal{U}$ biased toward lower values of \(t\), 
allocating more training budget near the data endpoint~$(t=0)$:
\begin{equation}
t \sim w\,\mathcal{U}(0,0.35) + (1-w)\,\mathcal{U}(0.35,1),
\end{equation}
where $w$ is the mixture weight, set to 0.7 to better capture high-frequency details. The training data consists of discrete samples \((c, \omega_o)\) obtained from the microgeometry sampling process. 
Additional data generation and training details are provided in the supplementary material.}

Once the flow matching network is trained, it can be used for outgoing direction sampling and pdf evaluation. For sampling, we use Euler integration with a constant step size to solve the ordinary differential equation~(ODE) that transforms samples from the base distribution into the final outgoing direction.


For the pdf evaluation, we run the flow backwards and compute the Jacobian determinant by computing the network gradients at each step of the ODE solver and accumulating them, following Fu et al.~\shortcite{Fu:2024:BSDFdiffusion}. We compute the final pdf by multiplying the pdf of the base Gaussian distribution by the accumulated Jacobian determinant. Since our pdf $\rho$ is defined in the projected solid angle measure, we transform the pdf value into the solid angle measure by multiplying by a cosine term.

\subsection{MLP for the view-dependent albedo term}
\label{sec:albedo}

The view-dependent albedo term maps the incoming direction $\omega_i$ into an RGB value. As it exhibits smooth behavior, we use a small MLP to learn it, consisting of two hidden layers, each with 32 neurons. We encode $\omega_i$ using fourth-order spherical harmonics.  We train the network with the $\mathcal{L}_1$ loss between the prediction and the Monte Carlo estimate of the albedo term for a given incoming direction. {More training details are shown in the supplementary.}



\subsection{Neural texture for spatially varying materials}
\label{sec:sv}

To add spatial variation capability to our PureSample framework, we introduce a neural (latent) texture. The latent code from each texel of this texture serves as an additional condition for flow matching, together with the previous conditions ($\omega_i$ and channel) to generate the distribution. We increase the number of neurons per layer to 128 for both networks, extend the albedo network to five layers, and add one residual layer, due to the increased complexity of spatially varying materials. 

For the loss function, we build upon Eqn.~(\ref{eq:FM_loss}) by adding an additional total variation (TV) loss to reduce noise in the neural texture: \begin{equation}
 \begin{aligned}
 L_{\mathrm{TV}}=\sum_{x,y}(\left|T(x+1,y)-T(x,y)\right|+\left|T(x,y+1)-T(x,y)\right|), 
 \end{aligned} 
\end{equation} 
where $T(x,y)$ is the latent code in the neural texture at texel $(x,y)$.

We first train the flow matching network to obtain the neural texture. For this, we randomly select locations from the neural texture, generate refined discrete samples, and use these samples to train the flow matching network while updating the neural texture. Next, we train the albedo network with a frozen neural texture.

Thanks to the neural texture, our representation can be applied to various spatially varying properties, such as microgeometries, colors, or normals, as illustrated in Figs.~\ref{fig:res_microstructure} and \ref{fig:gallery}.


\subsection{Rendering with neural PureSample representation}
\label{sec:rendering}

To use our PureSample representation for rendering, we need to support three key components: BRDF evaluation, importance sampling and pdf evaluation for multiple importance sampling (MIS). 

For BRDF evaluation, given incoming direction \(\omega_i\) and outgoing direction \(\omega_o\), we employ the flow matching network to obtain the pdf as the probability distribution term for each color channel (thus the network is inference three times in a batch). Subsequently, we query the albedo network to get the view-dependent albedo term. The product of these components yields the final BRDF value. 


For importance sampling, given the incoming direction \(\omega_i\), we first query the albedo network to acquire the view-dependent albedo term, \added{and then use Russian roulette to select a color channel, with the selection probability proportional to the albedo values.} We then use the flow matching network, conditioned on this chosen channel, to sample an outgoing direction. \added{For MIS, the sampling pdf is given by a three-channel mixture predicted by the flow matching network. Since both the BRDF evaluation and the MIS pdf require the same per-channel probability distribution term evaluated at identical $\omega_i$ and $\omega_o$, we reuse the network outputs across the two computations. }



\subsection{\added{Acceleration by MeanFlow and pdf distillation}}
\label{sec:acceleration}
\added{
As flow matching requires multi-step ODE integration for sampling, its low efficiency becomes severe when evaluating the pdf, as the additional Jacobian computation further increases the computational cost, limiting its applicability for rendering. As an optional acceleration strategy, we propose employing one-step sampling with MeanFlow~\cite{geng:2025:meanflow} and, independently, distillation pdf evaluation used in BRDF evaluation and MIS.
}

\added{
\paragraph{MeanFlow} We adopt MeanFlow as a replacement for flow matching to accelerate the sampling. MeanFlow generates samples in a single network forward pass, and its pdf evaluation requires a multi-step procedure. We leverage MeanFlow to accelerate sampling and accordingly modify the training and inference procedures.
}

\added{
\paragraph{Pdf distillation} 
We reduce the number of pdf evaluations of the neural sampling model to further accelerate rendering. Currently, our method for BRDF evaluation and MIS requires evaluating the pdf for each color channel independently. To reduce this cost, we approximate the pdf using an additional MLP when computing BRDF values and MIS weights. Meanwhile, we retain the exact pdf evaluation using the neural sampling model for unbiased BRDF sampling.
}


\added{With these accelerations, our method achieves approximately an $8\times$ speedup with only a slight degradation in quality. We therefore provide two variants: one optimized for high visual quality (without acceleration) and another optimized for efficiency (with acceleration enabled). Notably, our method is not intrinsically tied to any specific neural sampling model. We further demonstrate the applicability of our approach using normalizing flows, with implementation details provided in the supplementary materials.}



\section{Results}
\label{sec:result}


We implement rendering within the Mitsuba3 renderer~\cite{jakob2022mitsuba3} with CUDA enabled. The training and rendering are performed on an NVIDIA RTX 4090 GPU. The resolution of all renderings is set to $512\times512$.

\subsection{Materials with microgeometries}


\added{We validate our method on different microgeometries. As ground truth (GT), we render the explicit meshes generated from these microgeometries using path tracing with 1024 samples per pixel (spp). We also render our PureSample representation on both a flat quad and the BentQuad scene, under four point light sources and an environment light. \revise{All results use our SV variant.}}

\begin{figure*}[tbp]  
\centering
\includegraphics[width=\linewidth]{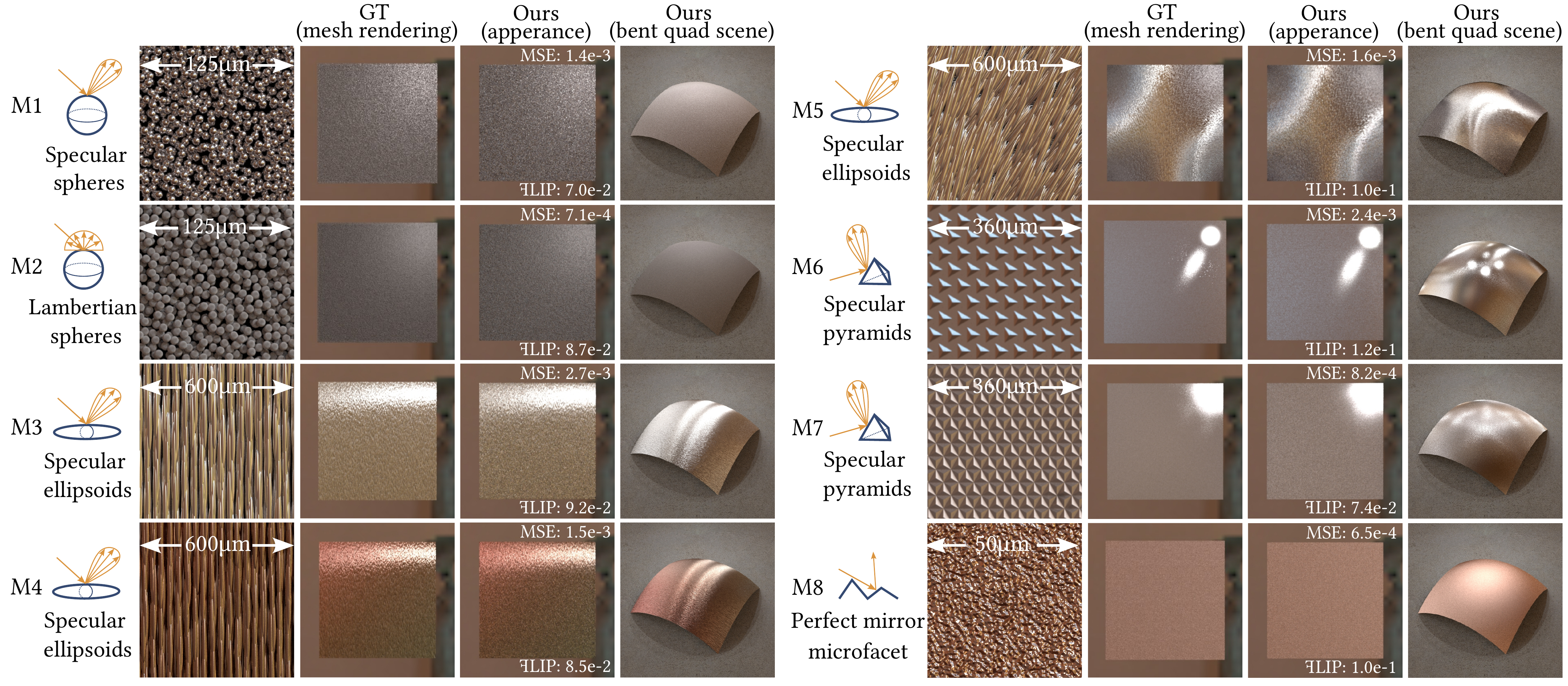}
\caption{\added{Validation of our PureSample on various microgeometries, where the microgeometry and micro-BRDF are shown on the left. The ground truth is rendered using path tracing on microgeometry meshes, while our method acts as an appearance model on a quad. Our method produces renderings that closely match the ground truth over various microgeometries, particularly demonstrating anisotropic highlights and a color shift. Assuming the BentQuad is 10 cm × 10 cm, the dimensions of each microgeometry are annotated in the figure.}}
\label{fig:res_microstructure}
\end{figure*}

\begin{figure*}[tbp]  
\centering
\includegraphics[width=1.0\linewidth]{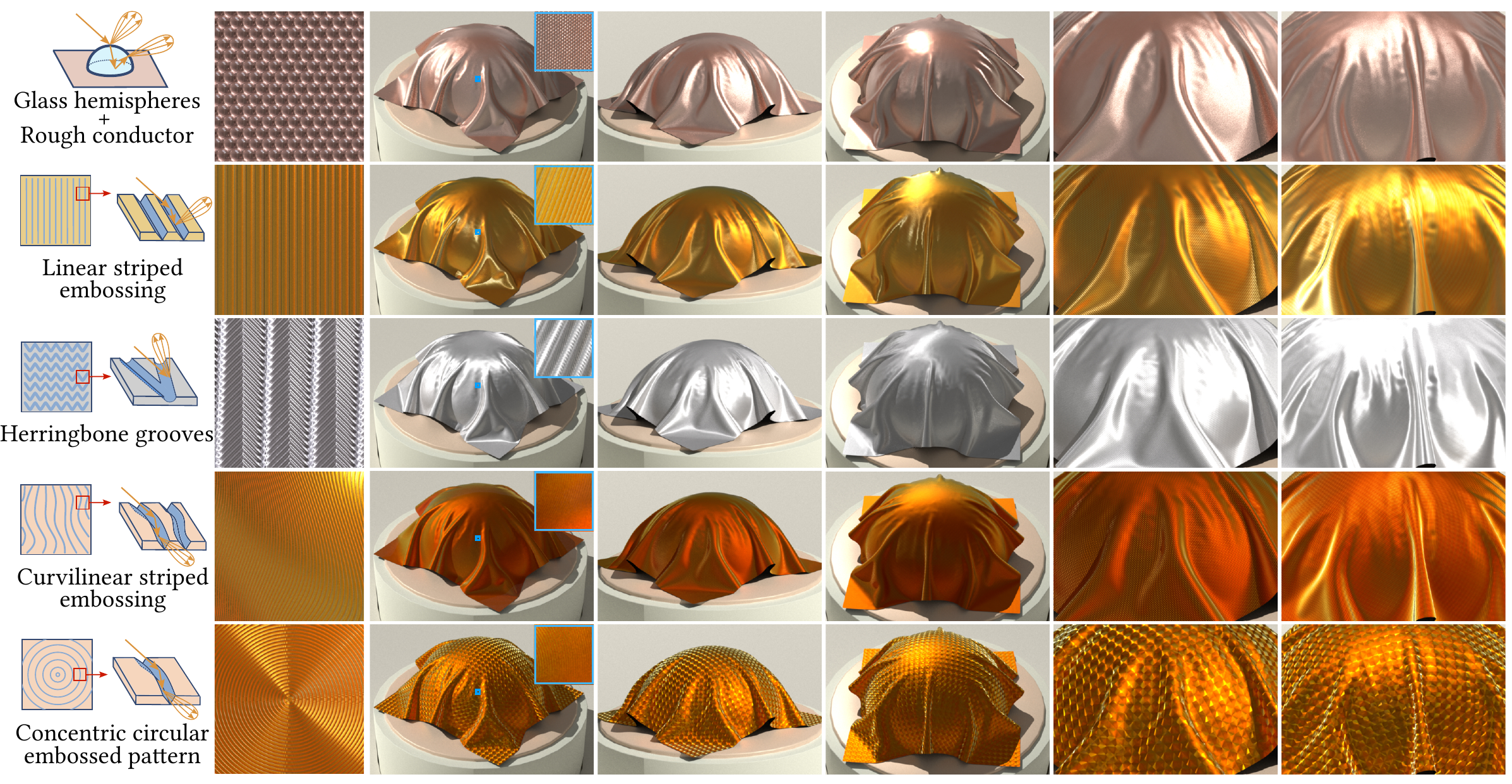}
\caption{\added{A gallery of diverse microgeometry materials represented by our method (with acceleration enabled). We render these microgeometries as surface appearances on a draped cloth under both environment and directional lighting, and the zoom-in results are shown in the blue box at the first view. Please refer to the supplementary video for animated results, which are important for observing the characteristic specular effects produced by the microgeometries.} \revise{Assuming a fabric size of $20\mathrm{cm} \times 20\mathrm{cm}$, the corresponding micro-renderings are at a scale of $40\mathrm{\mu m}$, except for the last row. The last row uses a larger-scale material composed of repeating tiles (each $4\mathrm{mm}$), where each tile contains concentric circular embossings of size $10\mathrm{\mu m}$.}}
\label{fig:gallery}
\end{figure*}

\begin{figure}[tbp]
\centering
\includegraphics[width = 0.95\linewidth]{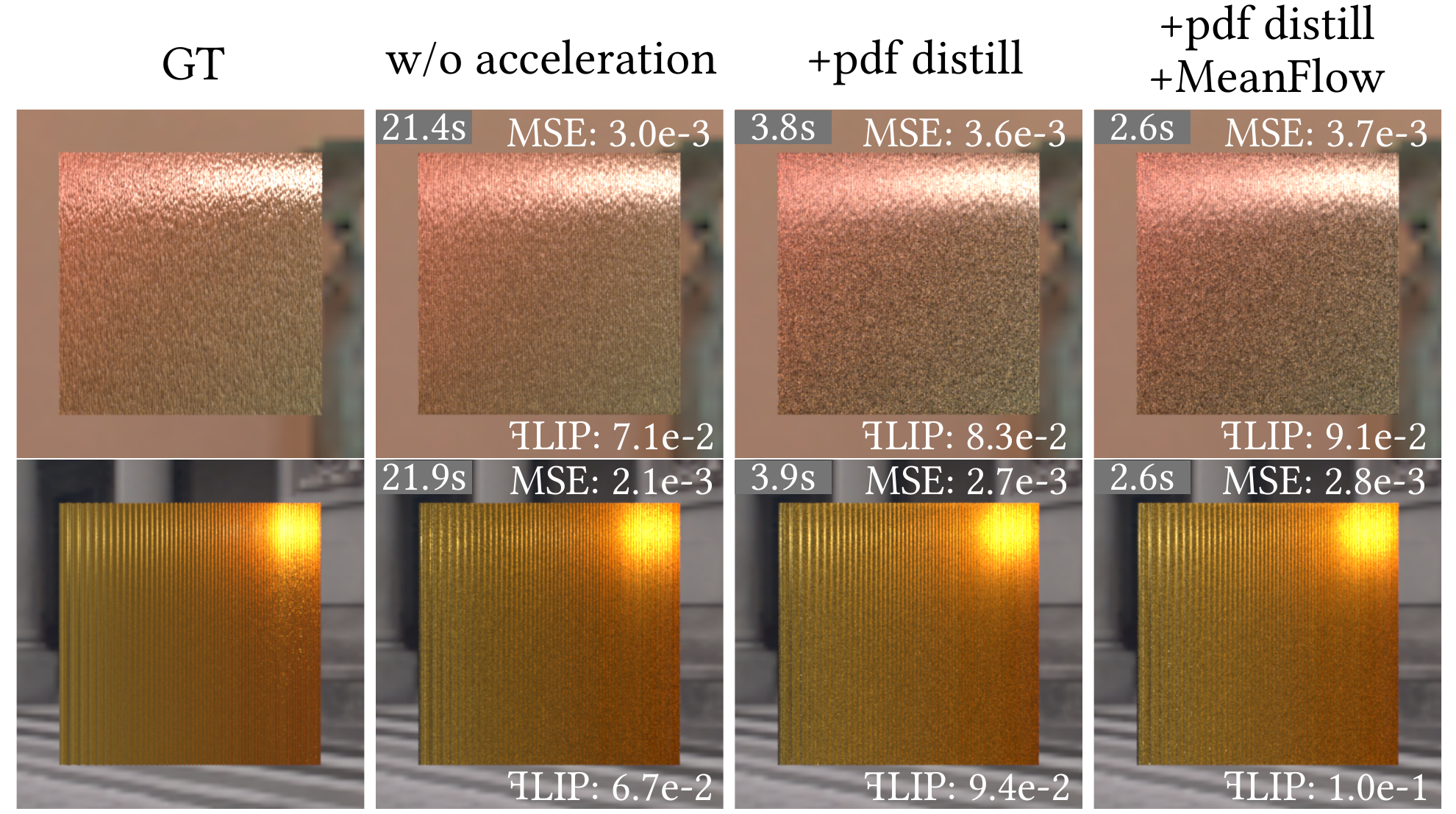}
\caption{\added{We compare our method before and after acceleration. Both pdf distillation and MeanFlow one-step sampling improve rendering speed without sacrificing too much visual quality. }}
\label{fig:speed_up}
\end{figure}

\added{
Fig.~\ref{fig:res_microstructure} presents results on a diverse set of microgeometries, including spherical, ellipsoidal, pyramidal, and microfacet-based structures, exhibiting a wide range of diffuse and specular appearances. Variations in micro-BRDF, orientation, distribution patterns, and spatially varying coloration lead to distinct highlight behaviors and anisotropic effects. Across all examples, our renderings closely match the GT images. \revise{Compared with the ground truth, our method achieves a rendering speed-up ranging from $7.7\times$ to $8.8\times$, while reducing storage by about $2\times$. Fig.~\ref{fig:speed_up} presents qualitative results demonstrating an $8\times$ speedup with only a slight degradation in quality.} Detailed rendering times and storage costs are provided in the supplementary materials.

}
\added{In Fig.~\ref{fig:teaser}, we present various materials in a single scene, including different types of microgeometries and layered materials. In Fig.~\ref{fig:gallery}, we present a gallery showcasing a variety of interesting microgeometry materials represented by our method with acceleration enabled. The microgeometries of these materials lead to distinctive appearances, demonstrating the expressive capability of PureSample.}

\subsection{Complex analytical materials}

We validate our method on complex analytical materials, including the multi-bounce microfacet model~\cite{Cui:2023:GMBBRDF} and the position-free layered material model~\cite{Guo:2018:layered}. \revise{All results use our non-SV variant.}

\begin{figure}[tbp]
\centering
\includegraphics[width = 0.88\linewidth]{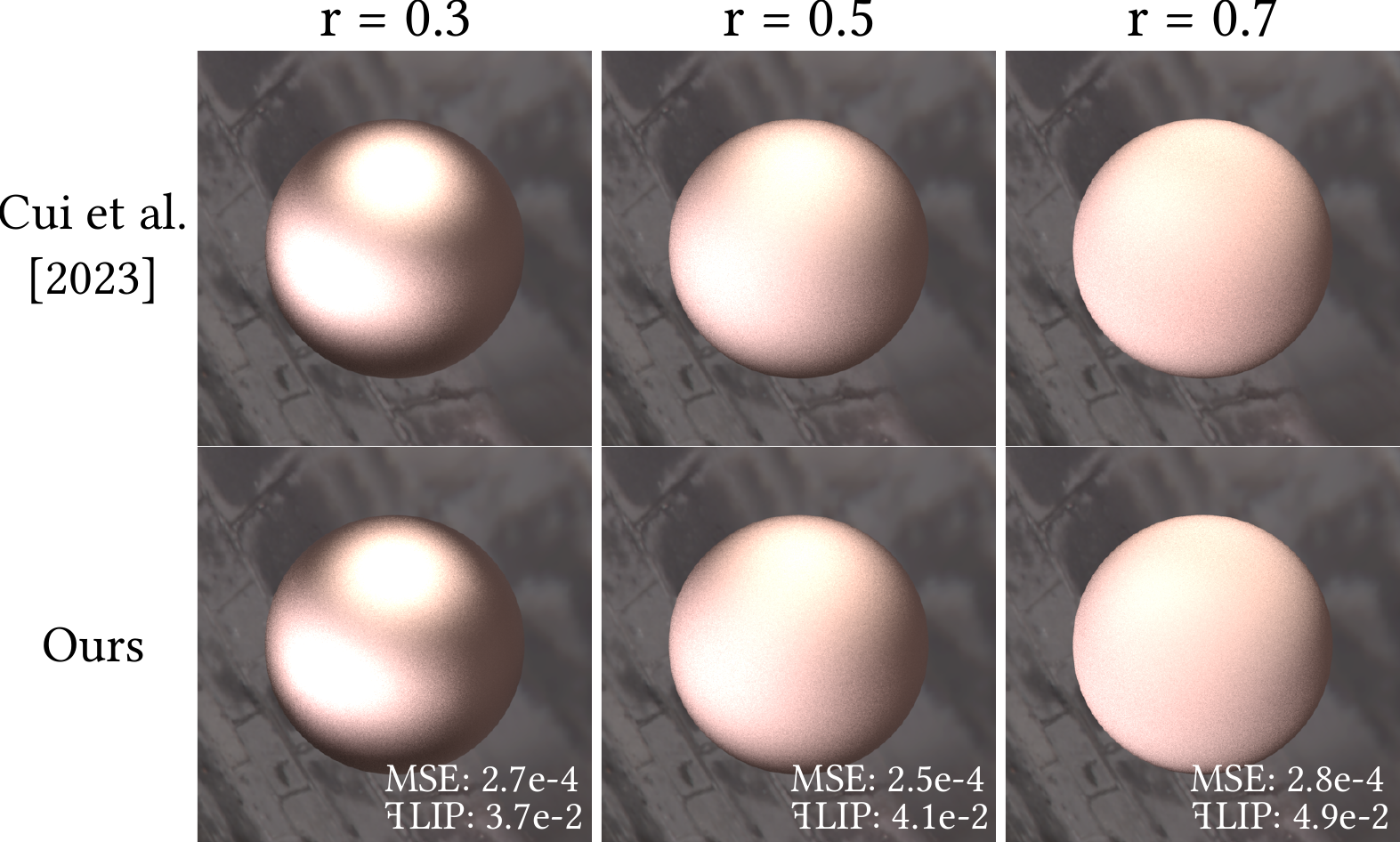}
\caption{Comparison of our multiple-scattering microfacet materials with Cui et al.~\shortcite{Cui:2023:GMBBRDF}. Our method produces results that closely match those of previous multi-bounce microfacet models over varying roughness values.}
\label{fig:res_mulscatteringWithCui}
\end{figure}

For the multi-bounce microfacet model, we generate the microgeometries procedurally using a Beckmann distribution. In Fig.~\ref{fig:res_mulscatteringWithCui}, we compare our PureSample against the previous work~\cite{Cui:2023:GMBBRDF}. Our method produces appearances that are consistent with the previous method across different roughness levels.

In Fig.~\ref{fig:res_layer}, we present both renderings and outgoing radiance distributions for various multi-layered materials. \added{For the data generation, we use forward sampling through layers, and do not use the NEE nor bidirectional estimators from previous work~\cite{Guo:2018:layered}.} Our approach produces renderings that closely match the ground truth across different material types, particularly those with anisotropic highlights. The outgoing radiance distribution aligns closely with the ground truth at the front view and grazing angles. \added{In the supplementary, we compare more outgoing radiance distributions with Guo et al.~\shortcite{Guo:2018:layered}, and we demonstrate our multi-layered results with spatial variation using normal maps.}


\subsection{\added{Comparisons and ablations}}

\begin{figure}[tbp]
\centering
\includegraphics[width =1.0\linewidth]{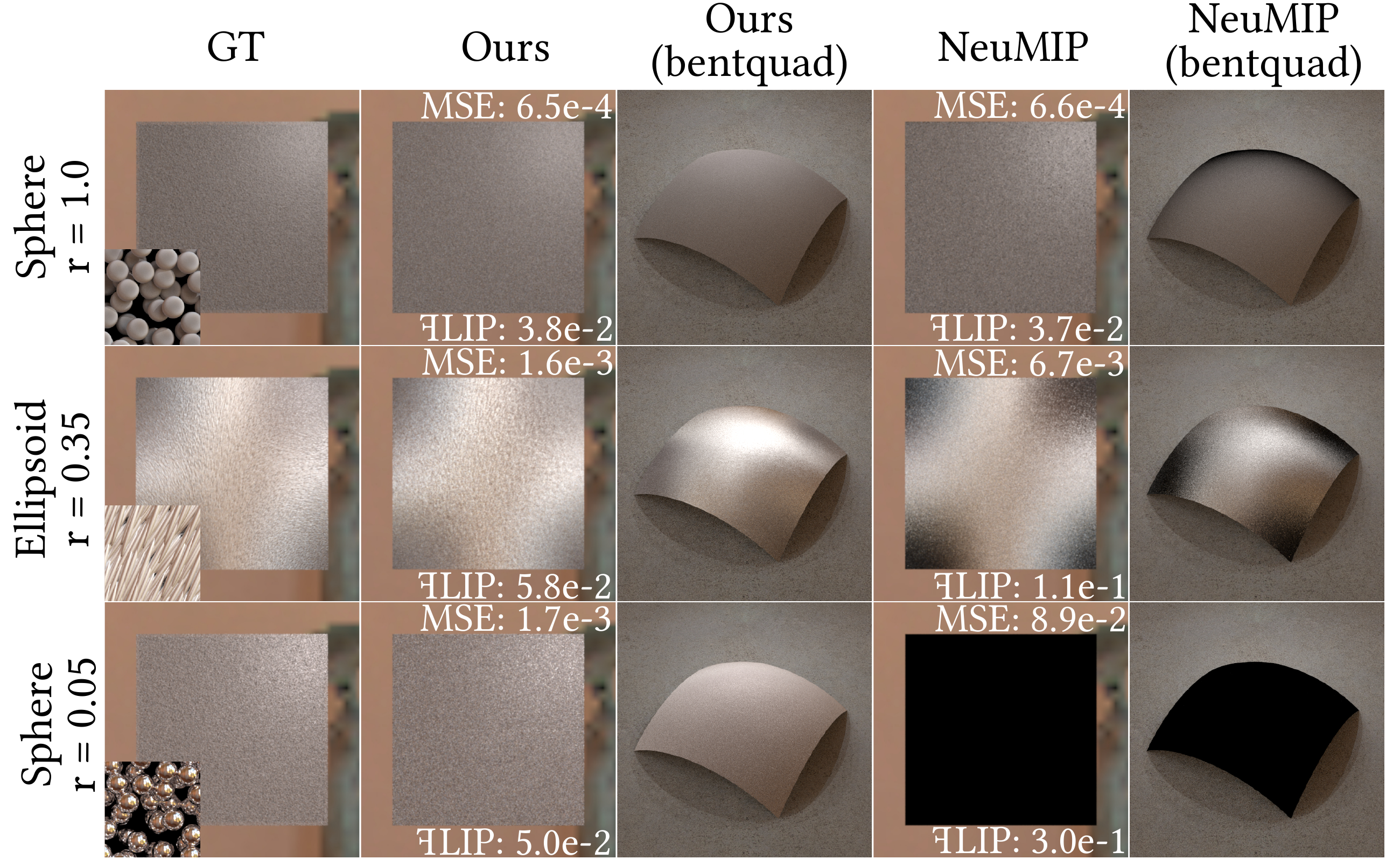}
\caption{\added{Comparison between our method and NeuMIP on materials with microgeometries of varying roughness. For microgeometries with lower roughness, NeuMIP exhibits noticeable artifacts or fails to produce valid results.}}
\label{fig:comp_neumip}
\end{figure}

\added{
\paragraph{Comparison to NeuMIP} We compare our method with NeuMIP \cite{Kuznetsov:2021:neumip} on materials with microgeometries.  As shown in Fig.~\ref{fig:comp_neumip}, NeuMIP is able to fit the appearance of microgeometry materials composed of diffuse micro-BRDFs. However, it produces noticeable artifacts for microgeometry materials with glossy micro-BRDF and completely fails on specular micro-BRDF materials. Following NeuMIP's original formulation, its training data is generated using NEE with a standard path tracer; this process introduces excessive noise in specular cases.
In addition, NeuMIP exhibits artifacts at grazing angles due to significant self-shadowing from microgeometry. Under such conditions, the NEE strategies used during path-traced data generation frequently fail, leading to a large number of zero-valued samples.
In contrast, our method constructs training data using a pure sampling strategy, thereby avoiding these issues. This further supports our observation that sampling is easier and more robust than direct BRDF evaluation.
}

\begin{figure}[tbp]  
\centering
\includegraphics[width=\linewidth]{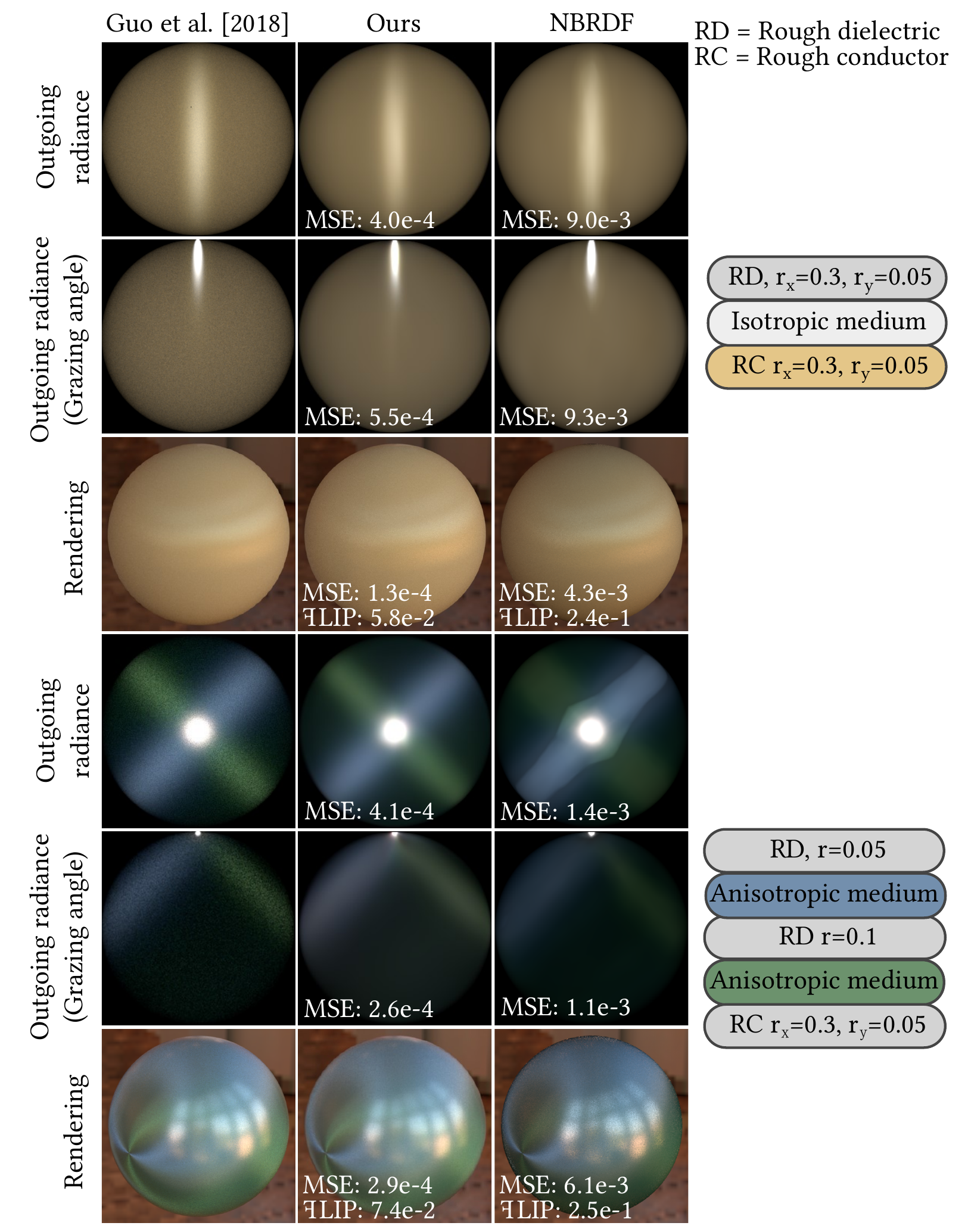}
\caption{\added{PureSample produces renderings and radiance distributions that closely match those of Guo et al.~\shortcite{Guo:2018:layered} on a wide range of multi-layered materials, including anisotropic double highlights. It achieves high-quality fitting at both grazing and non-grazing angles. Compared to NBRDF~\cite{sztrajman2021nbrdf}, our method achieves better fits.}}
\label{fig:res_layer}
\end{figure}

\added{
\paragraph{Comparison to NBRDF} We compare with NBRDF~\cite{sztrajman2021nbrdf} on layered materials. We extend the NBRDF network to match our network size for a fair comparison. Our training data is generated using pure sampling, whereas the data for NBRDF is produced via the much more complex bidirectional BRDF evaluation of Guo et al.~\shortcite{Guo:2018:layered}. As shown in Fig.~\ref{fig:res_layer}, our method achieves better fitting quality than NBRDF. BRDF data for layered materials generated via Monte Carlo simulation is inherently noisy, which poses challenges for the NBRDF network to handle, especially at grazing angles. In contrast, our flow matching network models the data as a distribution, enabling more effective learning from such noisy data.} 

\begin{figure}[tbp]
\centering
\includegraphics[width = 1.0\linewidth]{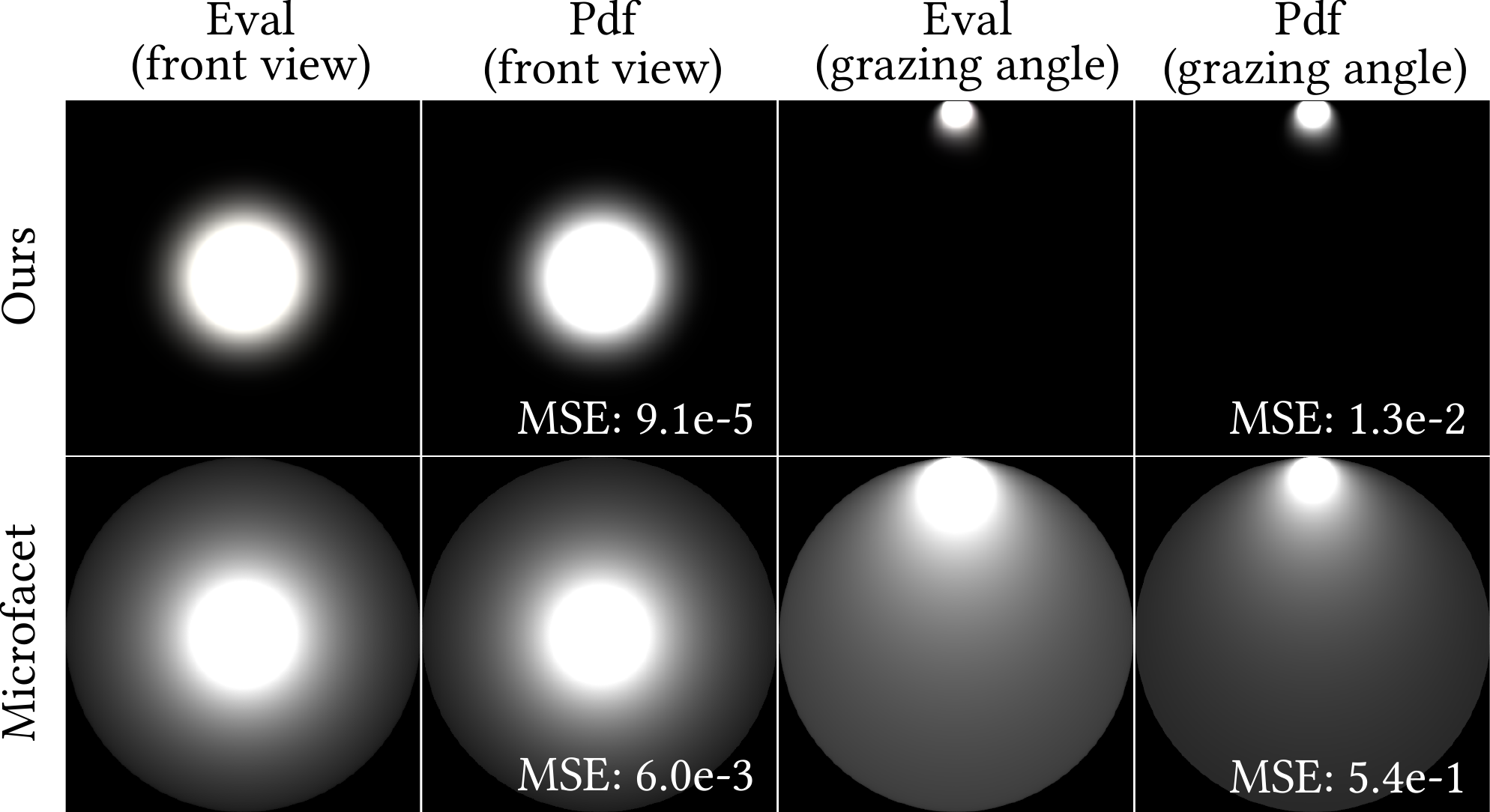}
\caption{\added{Compare our method with the analytical microfacet BRDF in terms of BRDF evaluation and importance sampling pdf. Compared to analytic microfacet materials, the importance sampling pdf of our method more closely matches the corresponding BRDF evaluation.}}
\label{fig:comp_anlypdf}
\end{figure}

\paragraph{\added{Comparison of eval\&IS pdf on ours and analytical BRDF}}
\added{We compare with the analytical microfacet BRDF in terms of evaluation / importance sampling match. As shown in Fig.~\ref{fig:comp_anlypdf}, compared to analytic microfacet materials, the importance sampling pdf of our method more closely matches its corresponding BRDF evaluation, since analytical importance sampling of microfacet BRDF (VNDF sampling) does not consider the Fresnel term. In contrast, our method learns perfect BRDF sampling by construction.}

\paragraph{Ablation} 
\added{Ablation studies on the TV regularization, neural sampling model, and network size are provided in the supplementary.}

\subsection{Discussion and limitations}
\label{sec:limitation}
\revise{We focus on representing BRDFs over the hemisphere, although the proposed method can be naturally extended to the full spherical domain to support BSDFs with transmission. Our current formulation targets microscale effects and does not account for mesoscale or macroscale parallax. Furthermore, changes to material properties require retraining the model, and our method does not currently support real-world data measurement, which we leave for future work. Additional discussion on the comparison with explicit microgeometry and the representation trade-offs is provided in the supplementary material.}
\section{Conclusion}
\label{sec:conclu}
We have introduced a new representation called PureSample, designed to model materials with arbitrary microgeometries or complex analytical material models. To achieve this, we decompose a BRDF into a probability distribution and view-dependent albedo terms, which are represented by a flow matching network and an MLP, respectively. Training these networks requires only forward particle tracing, eliminating the reliance on complex next-event estimation or bidirectional estimators. PureSample provides BRDF evaluation, importance sampling, and pdf evaluation, allowing for seamless integration into existing rendering frameworks. Through various material examples, we demonstrate the effectiveness and versatility of our method. As a novel BRDF representation, PureSample simplifies material design by removing the need for complex evaluation procedures and enabling the representation of any microgeometry at the surface appearance level. 





\begin{acks}
We thank the reviewers for the valuable comments. This work has been partially supported by the National Natural Science Foundation of China under grant No. 62572230.
\end{acks}
 

\bibliographystyle{ACM-Reference-Format}
\bibliography{bib}

\appendix
\begin{figure*}[!t]
\centering
\includegraphics[width = 1.0\linewidth]{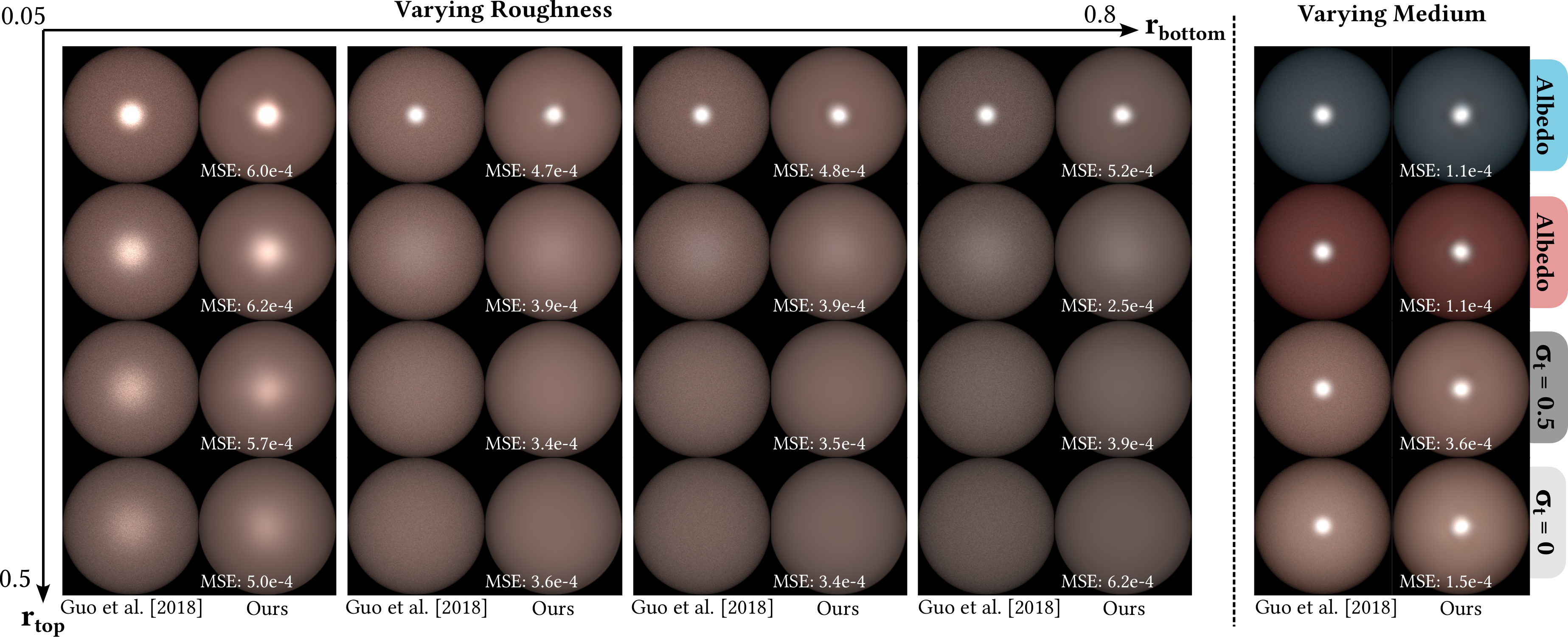}
\caption{Comparison of outgoing radiance distribution (incident elevation angle set to 0) between our model and Guo et al.~\shortcite{Guo:2018:layered} on varying roughness for both top and bottom layers and varying scattering albedos and $\sigma_t$ for the medium. The results of our PureSample representation closely match the references.}
\label{fig:res_layerVaringParam}
\end{figure*}

\section{Implementation details}
\label{sec:impl}
In this section, we describe the implementation details.

\subsection{Flow matching}
\label{impl:flow_matching}


\paragraph{Dataset generation}
We generate the training data using forward path tracing in Mitsuba~\cite{Mitsuba}. For each incident direction, we spawn rays from a plane placed sufficiently above the highest point of the microgeometry and trace them along the incident direction. This is always feasible for directions in the upper hemisphere (i.e., when $\cos\theta_i>0$), similar to NeuMIP~\cite{Kuznetsov:2021:neumip}. For grazing incidence, some rays may intersect the geometry before reaching the intended surface entry location. We interpret this as the genuine material behavior at grazing angles and learn the resulting particle distribution.

We use two types of strategies depending on the spatially varying property. For non-spatially varying materials, we uniformly sample $4,096$ $\omega_i$ over the hemisphere. For each $\omega_i$, we perform forward path tracing to generate $5,000$ outgoing directions ($\omega_o$) and compute the path ``throughput''. For spatially varying materials, we uniformly sample $500M$ pairs of $\omega_i,\omega_o$ over the hemisphere and compute the path ``throughput'' per pair. Data generation is conducted in parallel on an Intel i7-13700KF CPU. The time required for data generation varies according to the complexity of the material. For instance, generating data for a simple non-spatially varying two-layer material takes less than 30 seconds. In contrast, generating data for a more complex spatially varying material with microgeometries requires approximately 30 minutes.


\paragraph{Training}
We have implemented our flow matching network in PyTorch using the LAMPE library~\cite{tejero-cantero2020sbi}. The network is optimized for 500 epochs using the Adam optimizer with a learning rate of 0.003 to learn the target distribution. For spatially varying materials, we introduce a neural (latent) texture, which is jointly optimized with the flow matching network. The neural texture typically has a resolution of $256\times256$ and stores a 32-dimensional feature vector per pixel. The training is conducted on an NVIDIA RTX 4090 GPU, and the duration of training varies based on the complexity of the material. For instance, training a non-spatially varying two-layer material takes only five minutes, while training spatially varying materials with microgeometries requires up to 37.5 minutes. Tab.~\ref{tab:training_time} reports the per-material flow matching training time for the materials shown in Fig. 5 of the main paper.


\subsection{Albedo network}

\paragraph{Dataset generation}
\label{impl:albedo_MLP}

Following the same data generation strategy used for our flow matching network, we adopt two approaches depending on material complexity. For non-spatially varying materials, we directly reuse the dataset described in Sec.~\ref{impl:flow_matching}. For spatially varying materials, it differs from the flow matching because the view-dependent albedo term needs an expected value of the fraction of accepted samples. We uniformly sample 1M $\omega_i$ over the hemisphere. For each $\omega_i$, we also randomly sample a uv coordinate on the 0–1 plane and perform forward path tracing to generate 100 outgoing directions ($\omega_o$). Using the same hardware setup as described in Sec.~\ref{impl:flow_matching}, the dataset for our complex spatially varying materials with microgeoetries can also be generated within five minutes.

\paragraph{Training}
We have implemented our albedo network in PyTorch and optimize it using the Adam optimizer with a learning rate of 0.0001 for 3K iterations. For spatially-varying materials, we directly use the pre-trained neural texture values from Sec.~\ref{impl:flow_matching} as the conditioning input. On an NVIDIA RTX 4090 GPU, training the albedo network for a non-spatially varying two-layer material takes only 30 seconds, while the spatially varying materials with microgeoetries can be trained up to about two minutes. The albedo MLP training time of the materials in Fig. 5 (main paper) is also shown in Tab.~\ref{tab:training_time}.

\begin{table}
\centering
\caption{Training time per material corresponding to Fig.~5 (main paper). Our flow matching~(FM) network requires approximately 20–40 min to train, while the albedo network~(AN) converges in about 2 min. }
\label{tab:training_time}
\begin{tabular}{ccc|ccc} 
\toprule
\multicolumn{1}{l}{} & FM      & AN    & \multicolumn{1}{l}{} & FM      & AN     \\ 
\midrule
M1                   & 20.8m & 2.2m & M5                   & 34.0m  & 2.2m  \\
M2                   & 26.8m & 2.1m & M6                   & 37.5m & 2.2m  \\
M3                   & 34.1m & 2.1m & M7                   & 34.0m   & 2.1m  \\
M4                   & 30.3m & 2.1m & M8                   & 32.5m & 2.2m  \\
\bottomrule
\end{tabular}
\end{table}

\begin{table}
\centering
\caption{The rendering time (seconds) and storage cost (MB) of scenes in Fig.~5 in the main paper. Our method is treated as a surface appearance model on a quad and rendered with 16 spp, while the ground truth is rendered directly on the explicit microgeometry mesh with 1024 spp. Our method produces less noise and requires much fewer samples to converge. Our method requires less storage, and with pdf distillation and MeanFlow acceleration, achieves a substantial speedup over the ground truth. \added{In addition, Tab.~\ref{tab:time_breakdown} reports the rendering-time breakdown of different components before and after acceleration.} Note that the ground truth relies on a large collection of explicit microgeometries, making it difficult to render on arbitrary meshes. In contrast, our appearance model can be easily applied to any mesh.}
\label{tab:time_stor}
\begin{tabular}{ccccccc} 
\toprule
    & \multicolumn{2}{c}{GT} & \multicolumn{2}{c}{Ours} & \multicolumn{2}{c}{Ours (Acc.)}  \\ 
\midrule
    & Time  & Stor.~         & Time & Stor.             & Time       & Stor.               \\ 
\midrule
M1  & 23.5  & 25.4           & 21.9 & 8.6               & 2.6 (8.4$\times$) & 8.9                 \\
M2  & 17.9  & 25.4           & 21.7 & 8.6               & 2.7 (8.0$\times$) & 8.9                 \\
M3  & 74.2  & 18.7           & 21.8 & 8.6               & 2.5 (8.7$\times$) & 8.9                 \\
M4~ & 213.9 & 34.0           & 22.1 & 8.6               & 2.5 (8.8$\times$) & 8.9                 \\
M5  & 66.0  & 22.3           & 22.1 & 8.6               & 2.6 (8.5$\times$) & 8.9                 \\
M6  & 24.2  & 22.7           & 21.9 & 8.6               & 2.5 (8.7$\times$) & 8.9                 \\
M7  & 50.6  & 17.2           & 21.7 & 8.6               & 2.8 (7.7$\times$) & 8.9                 \\
M8  & 12.6  & 160.0          & 21.7 & 8.6               & 2.5 (8.6$\times$) & 8.9                 \\
\bottomrule
\end{tabular}
\end{table}

\begin{table}
\centering
\caption{\added{Average rendering time breakdown per spp for the flat quad rendering. With acceleration enabled, our method achieves nearly an $8\times$ speedup. See Fig.~7 in the main paper for an ablation of the acceleration strategies.}}
\label{tab:time_breakdown}
\begin{tabular}{ccc} 
\toprule
\multicolumn{1}{l}{} & sample & evaluation\&MIS pdf  \\ 
\midrule
w/o Acc.               & 0.728s & 0.579s             \\
w/ Acc.                 & 0.134s & 0.034s             \\
\bottomrule
\end{tabular}
\end{table}

\subsection{Acceleration strategies}

\paragraph{MeanFlow} MeanFlow parameterizes average velocity by a network between time $t$ and $r$, its objective is as below:
\begin{equation}
    \begin{aligned}
    \mathcal{L}_{\text{MF}}(\theta)&=\big\| u_{\theta}(x_{t}, c, r,t)-\text{sg}(u_{\mathsf{tgt}})\big\|^{2}, \\
    u_{\mathsf{tgt}}&=(x_1-x_0)-(t-r)\text{JVP}(u_{\theta};x_1-x_0),
    \end{aligned}
\end{equation}
where $\text{JVP}$ is the Jacobian-vector product and $\text{sg}$ denotes stop-gradient, which helps create an apparent target for training.
The data generation, network architecture, and training procedure follow our flow matching setup. To enable pdf evaluation, we set $t = r$, which degenerates the MeanFlow formulation to standard flow matching, and use the Euler method for integration.

\paragraph{Pdf distillation} For the pdf distillation, we use an MLP with five hidden layers, each containing 128 neurons. We train the network using a log-space $L_1$ loss. We generate training data $\omega_i$ and $\omega_o$ 
by uniformly sampling their half and difference vectors and evaluating the pdf in a online-fashion, with spatial positions uniformly sampled for spatially varying materials. We implement the network in PyTorch and optimize it with the Adam optimizer at a learning rate of 0.001 for 10K iterations, which takes approximately 20 minutes. We employ fourth-order spherical harmonics for $\omega_i$ and $\omega_o$.


\subsection{Rendering}

Our method can be easily integrated into any rendering framework. We have implemented it as a BSDF component within Mitsuba renderer 3 and perform rendering with path tracing. During rendering, evaluating our flow matching network requires solving an ODE. We use an Euler integrator, typically with 50 steps for most materials. 

When acceleration strategies are enabled, we replace MeanFlow with flow matching using a one-step for BRDF sampling. Meanwhile, for BRDF evaluation and MIS, we replace the pdf evaluation of the neural sampling model with a single MLP query.


Tab.~\ref{tab:time_stor} reports the rendering time and storage cost of our approach and path tracing baseline for the example shown in Fig.~5 of the main paper.
Our method exhibits similar time costs across these microgeometries, while path tracing varies due to the different number of bounces required for each microgeometry or the different number of triangles. After applying MeanFlow and pdf distillation acceleration, our method simplifies BRDF evaluation and MIS pdf computation to a single network call. BRDF sampling is also reduced to a single step using MeanFlow, resulting in approximately an $8\times$ speedup. In addition, our method requires less storage. Note that the ground truth rendering relies on a large collection of explicit microgeometries, making it difficult to apply to arbitrary meshes. In contrast, our approach models appearance directly and can be easily applied to arbitrary meshes for rendering, which constitutes a key advantage of our method.




\section{Results and ablation study}
\paragraph{Outgoing radiance distribution of multi-layered materials}
In Fig.~\ref{fig:res_layerVaringParam} we visualize the outgoing radiance distribution for a material consisting of a two-layer rough conductor with an intermediate isotropic medium, using an incoming elevation angle of 0 (i.e., aligned with the surface normal). Our results are compared against those from the method of Guo et al.~\cite{Guo:2018:layered}, which we use as ground truth. Across a range of parameter settings, our method closely matches the reference, demonstrating its effectiveness in capturing the appearance of complex layered materials.

\begin{figure}[t]
\centering
\includegraphics[width = 1.0\linewidth]{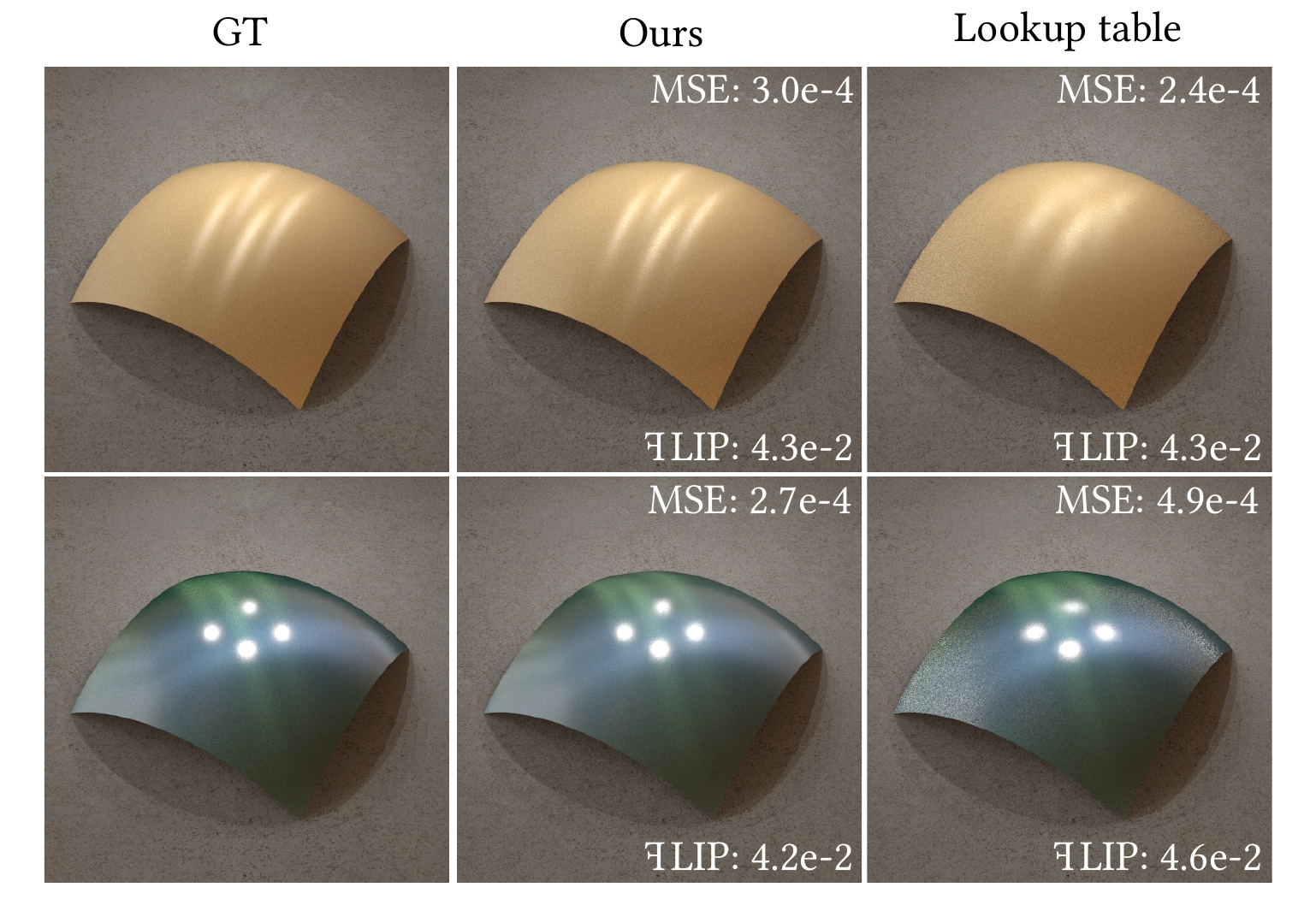}
\caption{\added{Comparison with a lookup-table approach~\cite{Ribardire:2019:NDFMicrogeo} on multilayered materials. The ground truth is rendered using Guo et al.~\shortcite{Guo:2018:layered}. The lookup-table method relies on a hemispherical sensor with limited resolution, leading to blurred highlights, whereas our method more accurately reconstructs specular details.}}
\label{fig:comp_tablet}
\end{figure}

\paragraph{\added{Comparison with lookup-table approach}}
\added{In Fig.~\ref{fig:comp_tablet}, we compare our method against a lookup-table approach~\cite{Ribardire:2019:NDFMicrogeo} on multilayered materials. The ground truth is generated using the method of Guo et al.~\shortcite{Guo:2018:layered}. We perform forward light tracing through the layers, and collect all outgoing radiance with a hemispherical sensor. Due to the limited resolution of the hemispherical sensor, our method produces a more accurate reconstruction of highlights. Although a higher resolution can improve the fitting quality, it also incurs a substantial storage cost, whereas our method is not constrained by this limitation.}

\begin{figure}[h]
\centering
\includegraphics[width=\linewidth]{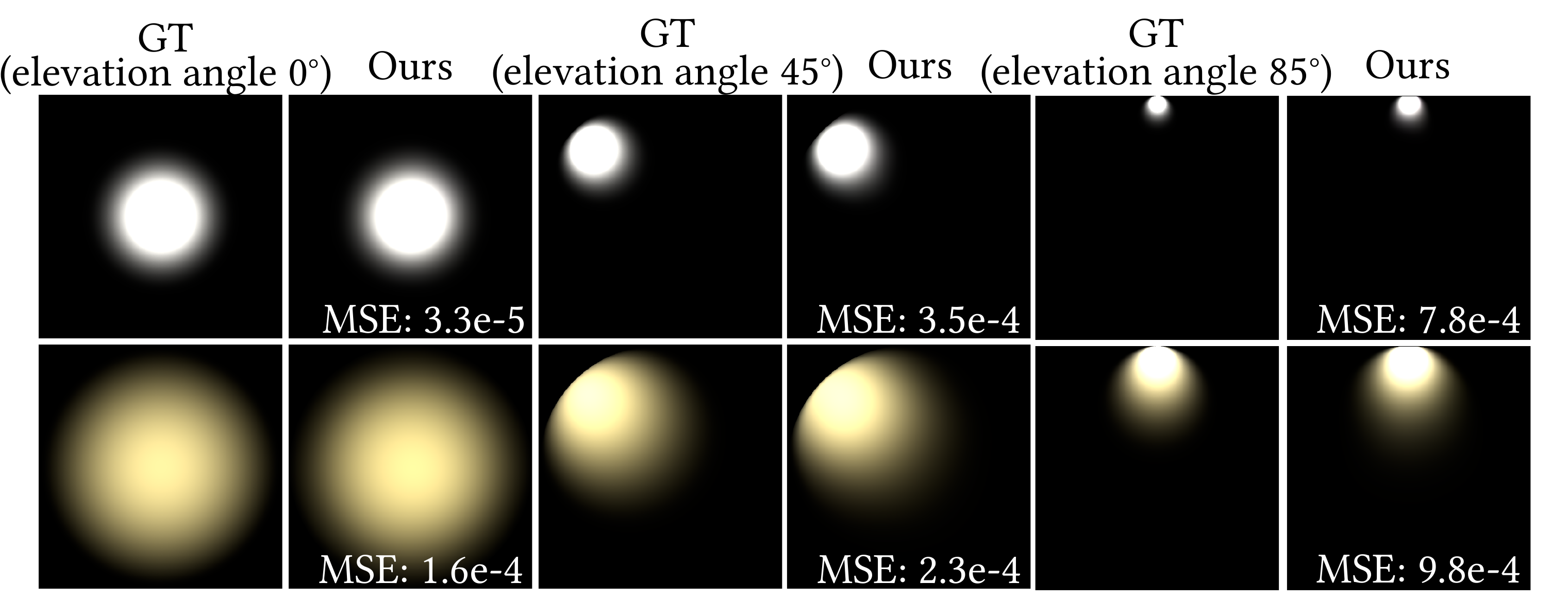}
\caption{\added{Validation on analytic microfacet materials. We compare 2D BRDF slices of the ground truth and our method for two materials under three incident elevation angles. Our results closely match the ground-truth slices across all cases.}}
\label{fig:comp_2Dslices}
\end{figure}

\paragraph{\added{Validation on analytic model}}
\added{We validate our method on analytic microfacet materials. In Fig.~\ref{fig:comp_2Dslices}, we show 2D BRDF slices of both the ground truth and our method for two microfacet materials under three different incident elevation angles. The first row corresponds to silver with a roughness of 0.1, and the second row corresponds to gold with a roughness of 0.25. Our results closely match the ground-truth BRDF slices across all evaluated directions.}

\begin{figure}[t]
\centering
\includegraphics[width = 1.0\linewidth]{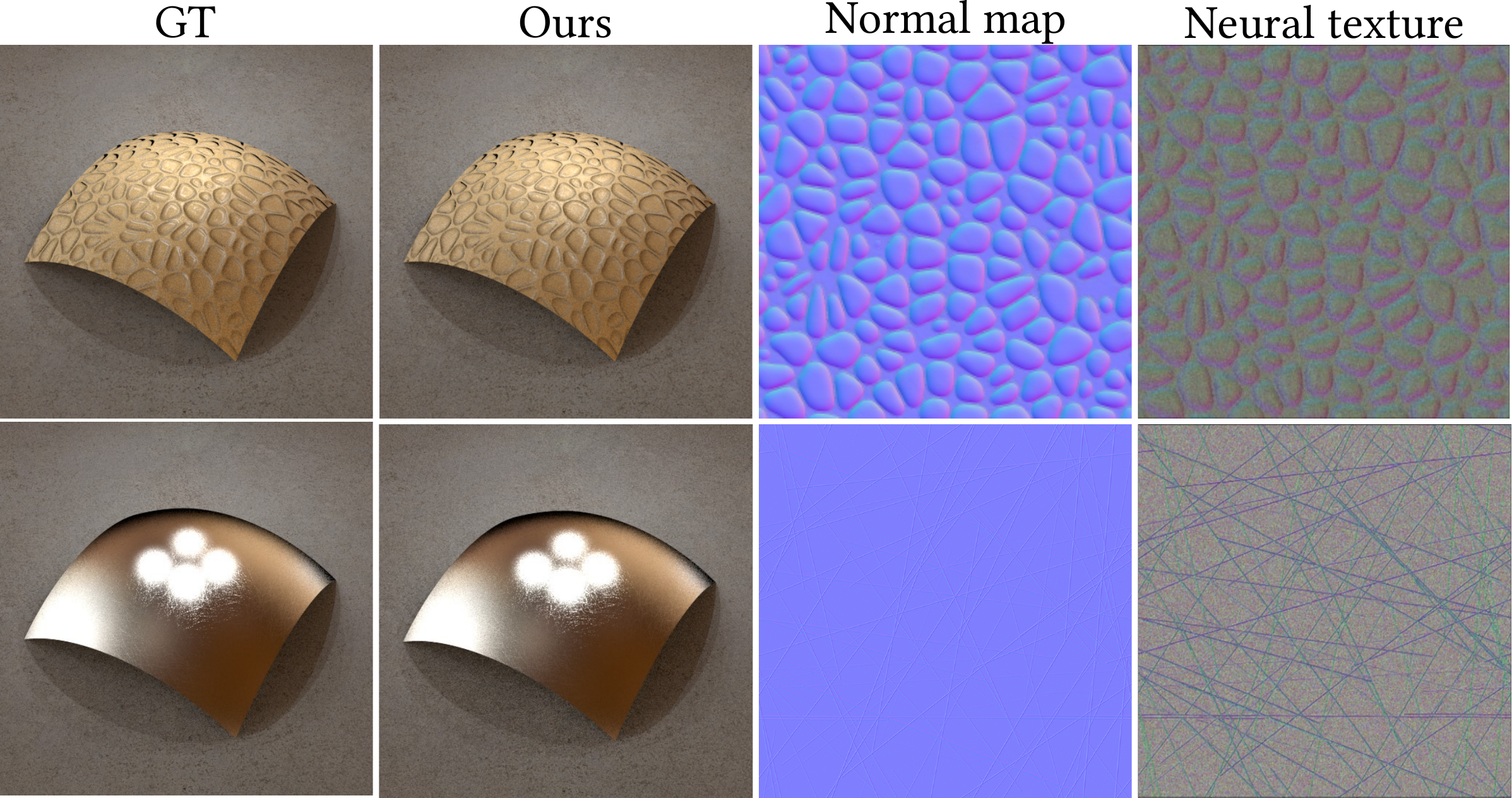}
\caption{Renderings and neural textures of spatially varying materials with our method. Our renderings are close to the ground truth, while the visualization of the neural texture reflects the normal variation.}
\label{fig:res_sv}
\end{figure}

\begin{figure}[t]
\centering
\includegraphics[width = 1.0\linewidth]{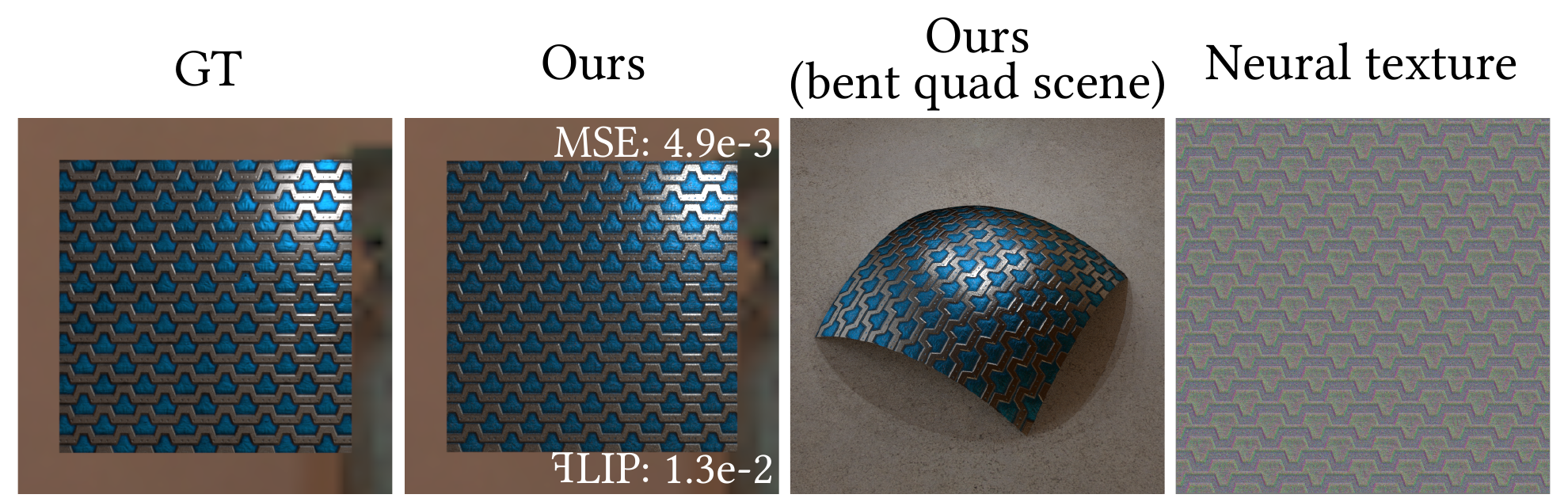}
\caption{\added{Complex SVBRDF based on a microfacet model with spatially varying roughness, normal, and IOR. Our method accurately reproduces the spatially varying appearance, closely matching the ground truth.}}
\label{fig:complex_sv}
\end{figure}

\paragraph{\added{Spatially varying material}}
\added{To demonstrate our method's capability with spatially varying materials, we conduct an experiment using materials with a normal map on the position-free layered material model~\cite{Guo:2018:layered}. As shown in Fig.~\ref{fig:res_sv}, our method generates renderings that closely resemble the ground truth, and the visualized neural textures also show a similarity to the input normal maps. We also show a complex SVBRDF based on a microfacet material in Fig.~\ref{fig:complex_sv}, where roughness and index of refraction (IOR) vary dramatically. Our method captures the complex spatially varying appearance consistent with the ground truth. }

\begin{figure}[h]
\centering
\includegraphics[width=.9\linewidth]{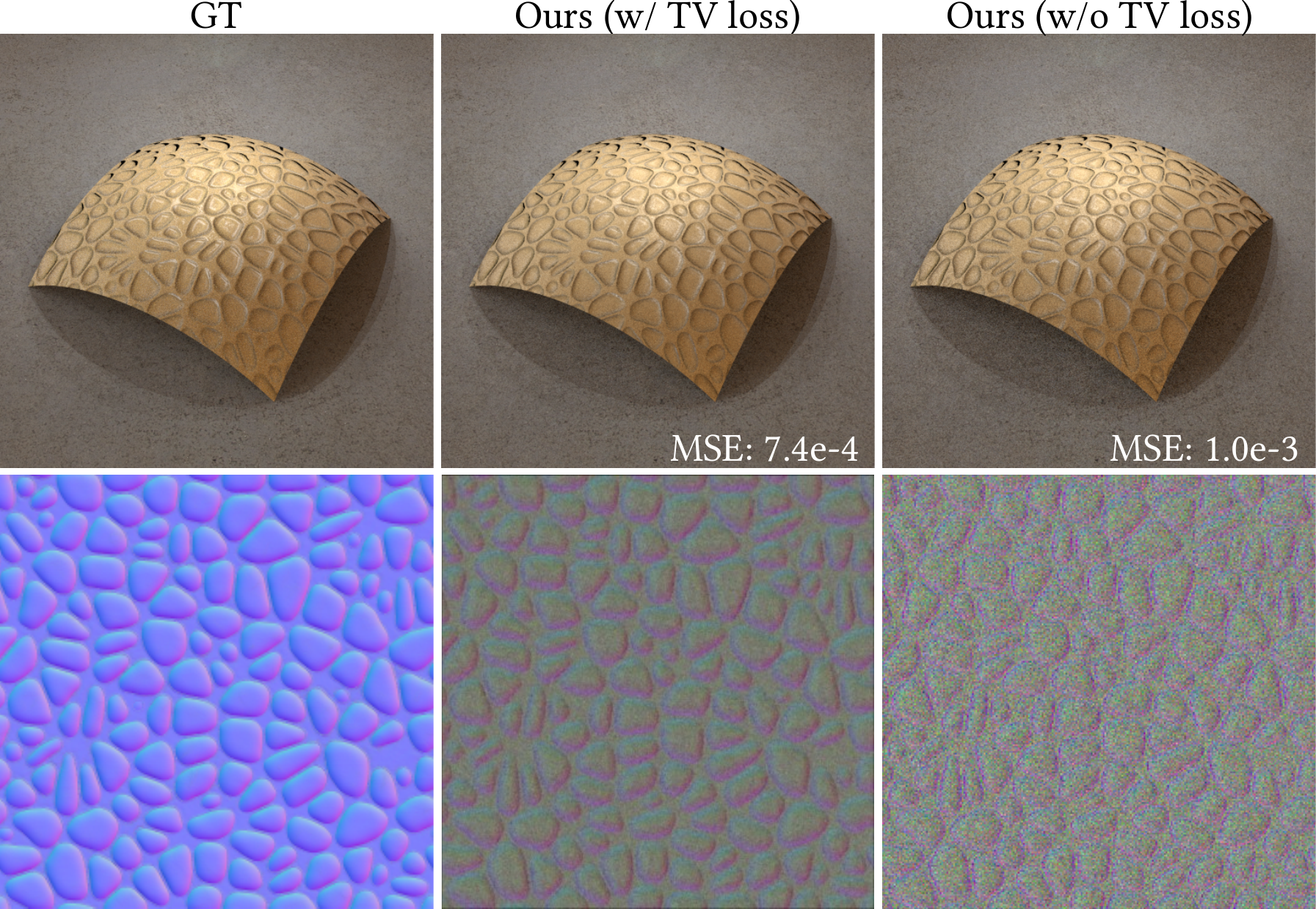}
\caption{Ablation of the neural texture TV loss. TV loss helps reduce the noise of our neural texture, leading to rendering results that are closer to the ground truth.}
\label{fig:ab_tvloss}
\end{figure}

\paragraph{Ablation of TV loss for the neural texture}
We evaluate the effect of the TV loss on our neural texture. As shown in Fig. \ref{fig:ab_tvloss}, the TV loss decreases noise in a multi-layered material with a normal map, resulting in an appearance that is closer to the ground truth.

\begin{figure}[t]
\centering
\includegraphics[width = .9\linewidth]{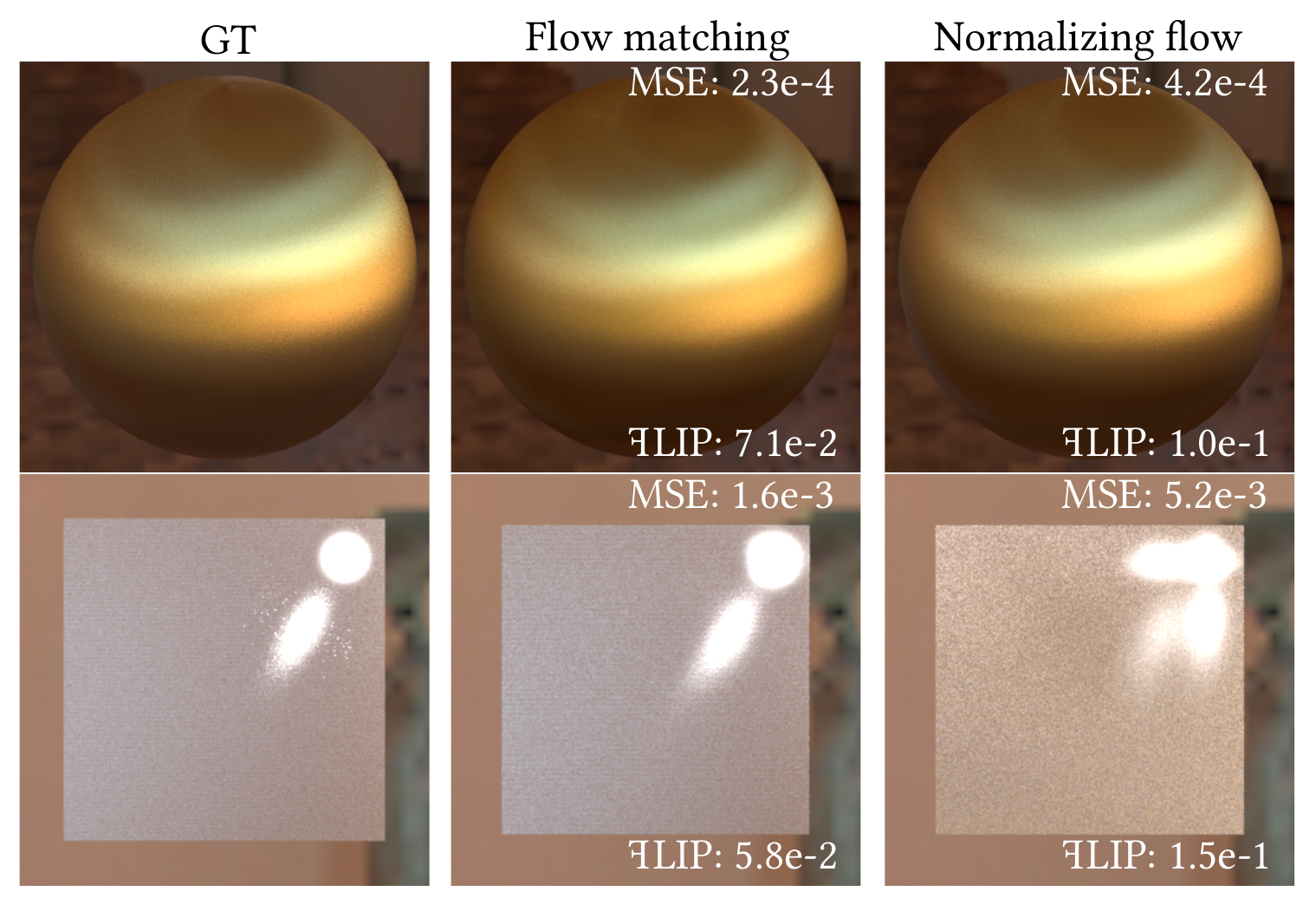}
\caption{We compare the results of using flow matching and normalizing flow as the neural sampling model on both layered material and microgeometry material. The normalizing flow uses a single coupling transformation and a network of the same size as that used for flow matching. For the layered material, both models perform well. However, for the microgeometry material, the normalizing flow is limited by its expressiveness and produces an incorrect appearance.}
\label{fig:comp_nf}
\end{figure}

\paragraph{Ablation on the choice of neural sampling model}
We demonstrate our method using different neural sampling models in Fig.~\ref{fig:comp_nf}, including flow matching and normalizing flows. For normalizing flows, we use Gaussianization flow~\cite{meng:2020:gaussianization} with a single coupling transformation and a network of the same size as that used for flow matching. While our approach is compatible with different model families, normalizing flows exhibit limited expressiveness compared to flow matching. Consequently, we select flow matching for sampling.

\begin{figure}[t]
\centering
\includegraphics[width = .9\linewidth]{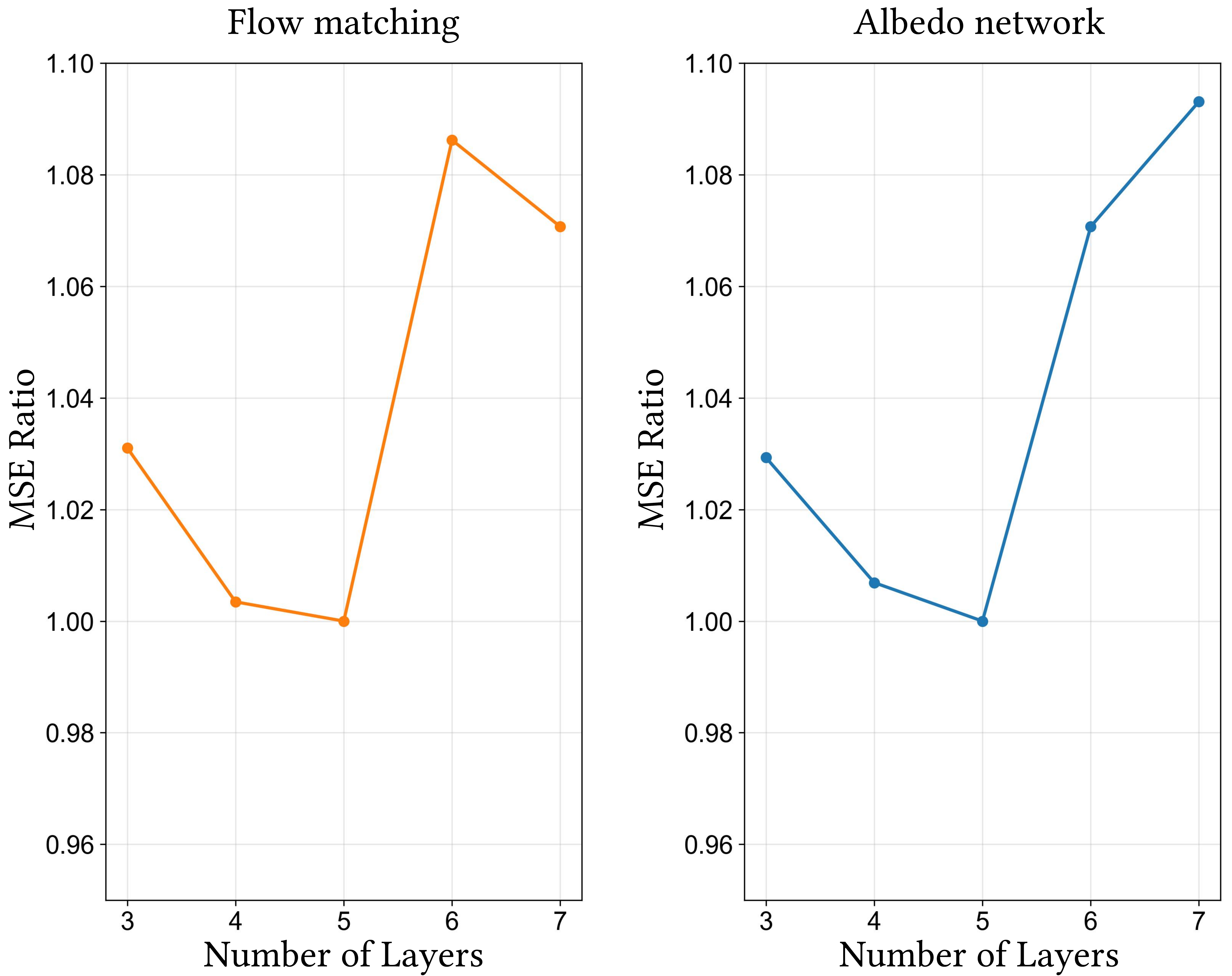}
\caption{Ablation of neural network size.
Left: the ratio of MSE errors of different network sizes relative to the default network size (5 layers) for the flow matching.
Right: the ratio of MSE errors of different network sizes relative to the default network size (5 layers) for the albedo network.}
\label{fig:comp_size}
\end{figure}

\paragraph{Ablation of network size}
We investigate the impact of different network architectures on result quality. 
Fig.~\ref{fig:comp_size} reports the ratio of MSE errors for networks with varying numbers of layers relative to the default architecture, for both the flow matching network and the albedo network. The current design achieves a good trade-off: using fewer layers results in higher errors, while larger networks increase training difficulty and do not yield further improvements.

\begin{figure}[t]
\centering
\includegraphics[width = 1.0\linewidth]{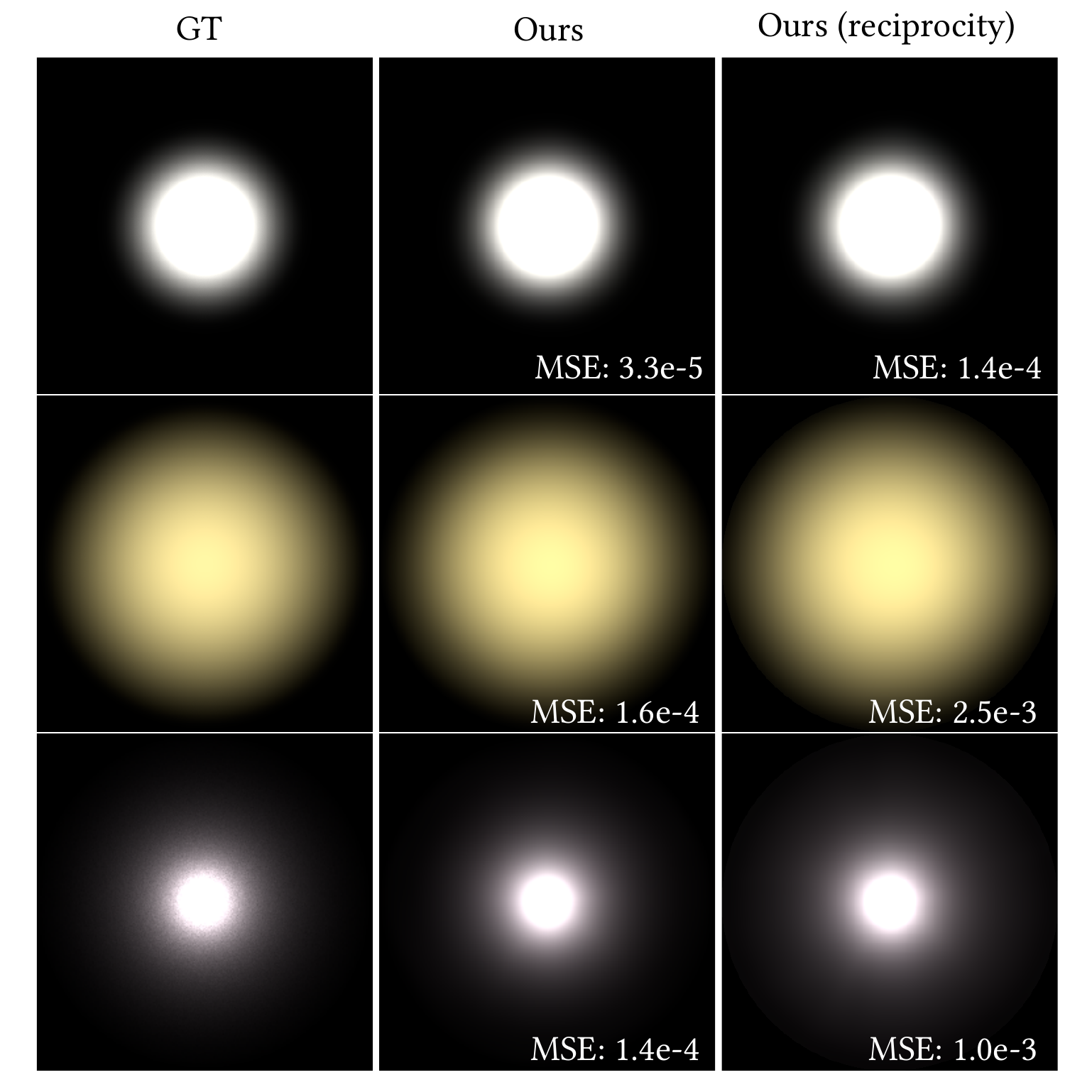}
\caption{\added{We compare the BRDF evaluation and its transpose (swapping $\omega_i$ and $\omega_o$) on the microfacet BRDF. Despite not explicitly enforcing reciprocity, the transposed BRDF evaluation remains visually consistent, indicating the limited practical impact of reciprocity violations.}}
\label{fig:reciprocity}
\end{figure}

\section{\added{Physical BRDF properties}}
\added{For physically based materials, the BRDF satisfies the properties of energy conservation, reciprocity, and pdf normalization. In this section, we discuss these three properties of our method.}

\paragraph{\added{Energy conservation}}
\added{Energy conservation in our model is governed by the view-dependent albedo term. We empirically evaluate this term across 10 materials with $10^8$ samples each, and observe that its value remains below 1 in all cases. In practice, we further enforce strict energy conservation by clamping the albedo term to the range $(0,1)$ during evaluation.}

\paragraph{\added{Reciprocity}}
\added{While reciprocity is not explicitly enforced in our formulation, which is consistent with prior neural material models (e.g., NeuMIP~\cite{Kuznetsov:2021:neumip}, NLBRDF~\cite{Fan:2022:Neurallayer}) and BTF-based representations. In practice, we observe that the transposed BRDF evaluation (swapping $\omega_i$ and $\omega_o$) remains visually plausible, with no noticeable artifacts (Fig.~\ref{fig:reciprocity}), suggesting that reciprocity violations are limited.}

\paragraph{\added{Pdf normalization}}
\added{Flow matching and normalizing flows transform a base Gaussian distribution into the target distribution. Since the base distribution is normalized, the target distribution remains normalized via the change-of-variables formula. In practice, a small fraction of samples may map outside the unit disk (i.e., the projected hemisphere). In our experiments (10 materials with $10^8$ samples each), only 0.045\% of samples fall outside, indicating a negligible impact. This issue could be further mitigated by adopting a slope-space parameterization.}

\section{\added{Discussion}}

\paragraph{\added{The benefits of our method compared to explicit microgeometries}}
\added{Although our method introduces an upfront training cost (on the order of tens of minutes per material), it provides a learned shading model that supports eval, sample, and pdf, enabling efficient importance sampling and integration into standard rendering pipelines. Unlike explicit microgeometry, which is easy to forward-sample but costly to evaluate under specific lighting and viewing configurations, our model directly captures the scattering behavior. Furthermore, it can be applied to arbitrary base meshes, avoiding the need to construct detailed geometry. In practice, the training cost can be amortized across repeated uses of the same material.}

\paragraph{\added{Artifacts and loss of detail}}
\added{Our method may exhibit minor visual artifacts or loss of high-frequency details (e.g., slightly blurred highlights or deviations in fine micro-structure). This behavior primarily stems from the finite resolution of the neural texture used to parameterize the material. Our model aims to match the ground-truth appearance up to the level of detail supported by this representation, rather than reproducing every microscopic feature exactly. Increasing the texture resolution can further improve fidelity, but introduces additional training time and storage costs. This reflects a trade-off between representation capacity and efficiency that is inherent to our design.}

\section{Validity of the Monte Carlo process}
In the main paper, we have defined a particle tracing Monte Carlo process, which given incoming direction $\omega_i$ traces a random walk and returns a sampled outgoing direction $\omega_o$ with a unit throughput, or nothing, achieved by Russian roulette rejection sampling. 

We would like to show that the distribution of ray directions $\omega_o$ leaving the surface is precisely $\rho(\omega_o | \omega_i)$, and the expected value of the fraction of accepted samples is precisely $\alpha(\omega_i)$.

Assume the microgeometry is lit by distant environment illumination $t(\omega)$. We can reuse the Monte Carlo process to compute an unbiased estimate of the reflected radiance $L(\omega_i)$ as follows:
\begin{itemize}
    \item Return $t(\omega_o)$ if the process returns a sampled direction $\omega_o$.
    \item Return 0 if the process terminated due to rejection.
\end{itemize}
The above is a standard path tracing estimator with Russian roulette and no next-event estimation, which is well known to be unbiased. Therefore, the above gives an unbiased estimate of the reflection for any test function $t$:
\[
\int_{\hemi} f(\omega_i, \omega) t(\omega) \dproj(\omega).
\]

By setting $t(\omega) = 1$ in the above, we immediately obtain the second claim, namely that $\alpha(\omega_i) = \int_{\hemi} f(\omega_i, \omega) \dproj(\omega)$ equals the expectation of the ratio of accepted samples.

For convenience, let us fix $\omega_i$ and drop it to simplify notation, and define the dot product of two hemispherical functions
\[
\langle a, b \rangle = \int_{\hemi} a(\omega) b(\omega) \dproj(\omega).
\]
In this notation, the estimator above is computing $\fdot{f}{t}$, where $f$ is the BRDF lobe.

To prove that the distribution of valid returned directions $\omega_o$ is indeed $\rho(\omega_o)$, let that distribution be $p(\omega_o)$; we just need to prove that $\fdot{p}{t} = \fdot{\rho}{t}$ for any test function $t$. 

However, $\rho$ is by definition equal to $f / \alpha$, so we need to show $\fdot{p}{t} = \frac 1 \alpha \fdot{f}{t}$. We already know that $\fdot{f}{t}$ is the expectation of the full estimator above. But $\fdot{p}{t}$ is simply the expectation of the modified estimator that does not return zero if a sample is rejected, but instead loops until a valid sample is found. The ratio of the values of these two estimators is clearly the fraction of valid samples, which we already proved is equal to $\alpha$.

\section{Sampling weights above 1}

Note, we are making the assumption that the micro-primitive importance sampling routines are good enough that none of their sampling weights will be above 1. If this is not the case, we have to clamp the weight to 1; this can be seen as modifying the microgeometry into a slightly less energy conserving one. In practice, this can always be avoided by using micro-primitives for which good analytic importance sampling is known; this is true for mirror reflections, Lambertian reflections and SGGX microflake distributions, among others.


\end{document}